\documentclass[journal=ancac3,manuscript=article]{achemso}
\usepackage{graphicx}
\usepackage{amsmath}
\usepackage{epsfig}
\usepackage{float}
\usepackage{lineno}
\usepackage{xcolor}

\title{Image-potential states on a 2D Gr-ferromagnet hybrid: enhancing spin and stacking sensing}
\author{Maciej Bazarnik}
\affiliation{Institute of Physics, University of Münster, D-48149 Münster, Germany}
\altaffiliation{Department of Physics, University of Hamburg, D-20355 Hamburg, Germany}
\author{Anika Schlenhoff}
\email{schlenhoff@uni-muenster.de}
\affiliation{Institute of Physics, University of Münster, D-48149 Münster, Germany}
\altaffiliation{Department of Physics, University of Hamburg, D-20355 Hamburg, Germany}

\begin{document}

\begin{abstract}
Spin-resolved scanning tunneling microscopy and spectroscopy studies of image-potential states on Gr/Fe/Ir(111) show their sensitivity to the spatial variation of the Gr-Fe distance, and of the interfacial charge and spin transfer within the moiré unit cell. 
A stacking contrast between fcc and hcp sites, indistinguishable in the direct tunneling mode, is provided by the image-potential states in both resonant tunneling topography and spectroscopy images. 
The spin-polarized measurements reveal a site- and energy-dependent spin-polarization of the image-potential states that is mapped across the moiré unit cell with a spin contrast between hcp and fcc sites unobservable in direct tunneling mode.
On the on-top sites, the lowest image-potential states are found to exhibit a high spin-polarization, in contrast to the electronic states around the Fermi energy evanescent into vacuum, and attributed to their interfacial character at the respective Gr-Fe distance.
Since the physisorbed graphene is only weakly spin-polarized on these sites, image-potential states reflect the spin density at buried interfaces covered by non-magnetic passivation layers. 
The simultaneous imaging of structural, electronic and magnetic properties, both at the surface and the buried interface will prove invaluable in research of 2D hybrids and heterostructures.
\end{abstract}
\maketitle
\section{Keywords}
Graphene, intercalation, STM/STS, image potential states, magnetism, buried interfaces, stacking contrast. 

\section{Introduction}
Image-potential states (IPSs) form a Rydberg-like series of unoccupied electronic states and resonances with a maximum probability density in front of the surface, vertically confined by the surface reflection properties and the collective response of the electron system that results in an attractive image charge~\cite{Echenique1978, Hoefer2016}. 
IPS properties can be studied locally utilizing scanning tunnneling microscopy (STM) and spectroscopy (STS) in the field emission mode~\cite{Binnig1985, Becker1985}.
Here, the electric field between the STM tip and the surface induces a Stark shift of the IPSs to higher energies potentially exceeding the local surface work function~\cite{Crampin2005}.
IPS wavefunctions span the entire vacuum gap between the STM tip and the sample, enabling resonant electron tunneling from the tip to the surface~\cite{Gundlach1966,Schlenhoff2020}.
By this means, IPSs have been used for revealing local surface properties unobtainable in the conventional direct tunneling mode of STM. 
Prominent examples are the atomic-scale imaging of insulating diamond~\cite{Bobrov2001}, the local analysis of the (spin-dependent) electron reflection at the surface~\cite{Caamano1999, Schlenhoff2019}, the identification of different metals on surfaces~\cite{Jung1995} or the resolution of atomic-scale spin textures at nm tip-sample-distances~\cite{Schlenhoff2020}.

With the increasing research interest in 2D materials, IPSs have regained attention~\cite{Borca2024}, known as sensitive probes for interfacial coupling and charge transfer at buried interfaces~\cite{Niesner2014, Bose2010, Liu2021}.
For graphene (Gr)-metal hybrids, the Gr distance to the metal is decisive for the degree of its hybridization and the corresponding charge transfer, but also determines the IPS properties~\cite{Borca2024, Armbrust2015, Armbrust2012, Borca2010}.
With decreasing distance, the double series of IPSs of freestanding Gr evolves into quantum-well like interfacial states interacting with both surfaces at intermediate distances of about $3.7~\mathring{A}$, and finally, at smaller distances of about $2.2~\mathring{A}$, into a single series of states that resemble those of a clean metal surface covered by a Gr spacer layer~\cite{Armbrust2015}. 
Deposed on a magnetic transition metal (TM), magnetism can arise in the Gr layer due to strong hybridization between Carbon~$\pi$ and TM~$d$ states~\cite{Weser2010,Brede2014,Decker2014,Decker2013,Vita2014}.
For chemisorbed Gr on Ni(111), conventional spin-polarized (SP)-STM achieved spin contrast, indicating that the electron density for states around the Fermi level $E_{\rm F}$ evanescent into vacuum is substantially spin-polarized~\cite{Dzemiantsova2011}.
Likewise, an exchange-split of the first IPS has been found by nonlinear photoemission experiments~\cite{Achilli2018}. 
On Ni(111), Gr grows as flat layer with a small Gr-Ni separation of about $2.1~\mathring{A}$~\cite{Voloshina2012}. 
In contrast, CVD grown Gr on Ir(111) exhibits a moiré superstructure resulting from the mismatch of the Gr lattice and the Ir substrate, the most commonly observed one being the R$0(10$x$10)$ on $(9$x$9)$~\cite{NDiaye2008}.
It is characterized by three high symmetry regions where the middle of the Carbon ring is located either on-top, in an fcc or hcp position over the substrate atom. 
TMs like Co or Fe grow pseudomorphically on Ir(111) under Gr and, therefore, the periodicity and orientation of the moiré is preserved~\cite{Bazarnik2015,Decker2014,Decker2013,Bazarnik2013}. 
However, while Gr interacts only weakly with Ir, hybridization with these TMs is stronger, and together with the lattice mismatch this typically leads to a pronounced buckling of the Gr layer.   
In case of Gr/Fe/Ir(111), the Gr-Fe distance changes from $3.6~\mathring{A}$ on-top to $2~\mathring{A}$ on the fcc site~\cite{Brede2014, Ferreira2018}, thus varying between a weakly interacting and a strongly interacting system within the moiré unit cell.
This varying hybridization leads to a modulation of the magnetic moments in the Fe, from $2.8\,\mu_B$ for atoms below Gr top sites down to $2.1\,\mu_B$ for fcc and hcp sites, accompanied by a considerable spin density transferred to Gr on the latter sites~\cite{Brede2014}. 
Moreover, compared to the pristine Fe/Ir(111), the exchange coupling J increases within the Fe layer and as a consequence this system changes from a skyrmionic into a ferromagnetic state~\cite{Brede2014,Sierda2019}.
Interestingly, the Gr-Fe distance variations in this system span exactly the range where drastic changes of IPS properties are expected~\cite{Armbrust2015}.

Moreover, while in conventional STM some systems show distinct moiré with easy to distinguish high symmetry areas~\cite{Decker2013,Bazarnik2013}, Fe-intercalated Gr exhibits only a strong contrast between the top area and the overall rest being therefore collectively called 'valley'~\cite{Decker2014}. 
Conventional SP-STM has revealed significant spin-polarization of the electron density around $E_{\rm F}$ on the valley sites, while the top sites exhibit only a weak polarization~\cite{Decker2014,Decker2013}, consistent with a spatially varying and periodic change of the magnetism in the Gr layer caused by the variation in structural geometry and hybridization within the moiré unit cell~\cite{Decker2014,Decker2013}. 
To the best of our knowledge, spin-polarized IPSs have been so far only studied on all-metal systems forming a single magnetic entity with a spatially uniform spin-polarization~\cite{Schlenhoff2020, Achilli2018, Schlenhoff2019, Schlenhoff2012, Kubetzka2007, Donath2007}. 
Proven to be sensitive to the charge transfer at buried Gr-metal interfaces, the question arises how IPSs do sense a respective spin transfer laterally varying within the moiré unit cell. 

Here, we report on a modification of the IPSs above this complex Gr-ferromagnet hybrid due to different local Gr registries within the moiré unit cell and a resulting high contrast between the fcc and hcp stacking regions in the resonant tunneling mode unachievable in conventional STM/STS.
We observe a moiré site- and energy-dependent effective spin-polarization of the IPSs that can be mapped across the entire moiré unit cell. 
Unlike the electronic states around $E_{\rm F}$, the lowest IPSs are found to exhibit a high spin-polarization on the on-top sites attributed to their quantum-well like interfacial character at the respective Gr-Fe distance.
This results in a previously unobserved increase in the spin-dependent signal when switching from direct to resonant tunneling at these sites.
Moreover, mediated by the IPSs, even a spin contrast between fcc and hcp sites of the moiré superstructure is observed. 

\section{Results and Discussion}
\begin{figure}[bt]
\begin{center}
\includegraphics[width=1\columnwidth]{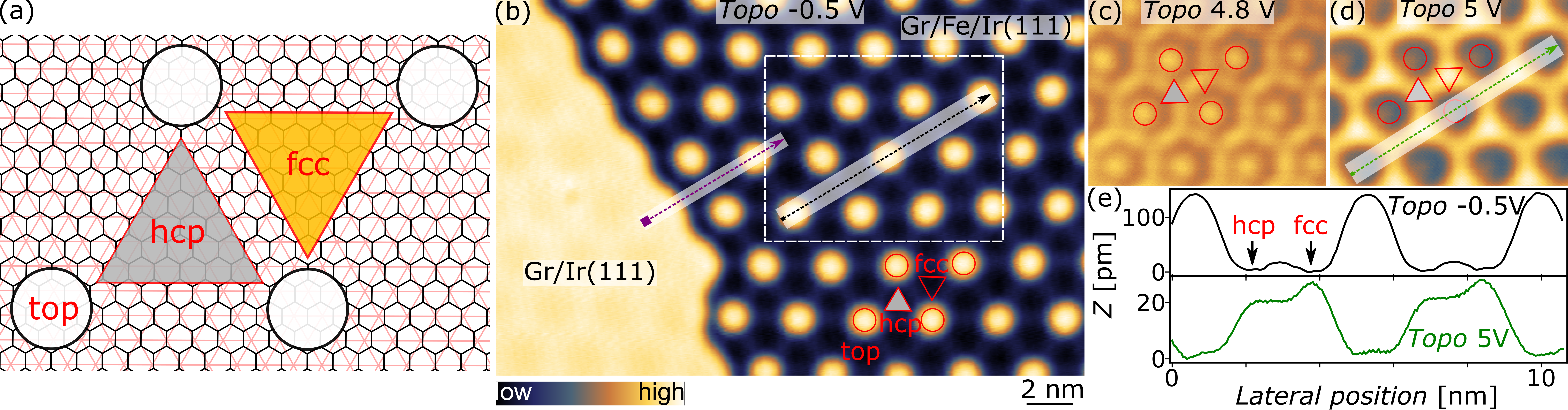}
\caption{\textbf{Comparison of stacking contrast in conventional and resonant tunneling topography mode.}
\textbf{(a)}~Model of Gr/Fe/Ir(111). Atoms lie at the intersections of each line. Red lines denote Fe and black lines denote carbon. Semitransparent circles mark top positions; a gray (orange) triangle denotes an hcp (fcc) region.
\textbf{(b)}~An overview conventional STM topography image recorded at $U=-0.5$\,V. Regions of Gr/Ir(111) are separated from Gr/Fe/Ir(111) by a buried Ir step edge. The black arrow indicates the line profile (black) shown in (e). The purple arrow indicates the direction of the line spectroscopy shown in Fig.~\ref{fig2}.
Resonant STM topography images recorded at \textbf{(c)}~$U=4.8$\,V and \textbf{(d)}~$U=5$\,V. While the apparent corrugation within moiré is substantially reduced compared to (b), top, hcp, and fcc regions (marked in red on all panels) are discernible in (d). The arrow marks the line profile (green) depicted in (e).
\textbf{(e)}~Apparent topography profiles for $U=-0.5$\,V and $U=5$\,V, respectively, with hcp and fcc regions marked. 
($I=1$\,nA).
Z range of the color scale in respective panels: (b)~$[0;185]~$pm, (c)~$[0;50]~$pm, (d)~$[0;50]~$pm.
}
\label{fig1}
\end{center}
\end{figure}
The model in Fig.~\ref{fig1}(a) presents the well-known structure of Fe-intercalated Gr on Ir(111)~\cite{Decker2014,Sierda2019}.
The size of the moiré is the same as for pristine Gr/Ir(111). 
In a conventional STM topography image, pristine and monolayer Fe-intercalated Gr on Ir(111) can be easily distinguished by a change in corrugation of the moiré pattern~\cite{Bazarnik2015,Brede2014,Decker2014,Bazarnik2013,Sierda2019} as observable in the overview image shown in Fig.~\ref{fig1}(b). 
Here, pristine Gr/Ir(111) and Gr/Fe/Ir(111) are separated by a buried Ir step edge, from where the Fe intercalation has started, growing on the lower terrace~\cite{Bazarnik2015}.  
The intercalated Gr shows a starlike moiré pattern with a pronounced corrugation about five times higher compared to that of pristine Gr~\cite{Bazarnik2015, Decker2014, Brede2014}.
The top sites are imaged as protrusions, whereas the fcc and hcp regions are separated by a small ridge of about $20$\,pm in height, as observable also in the line profile shown in Fig.~\ref{fig1}(e). 
The apparent height difference between fcc and hcp is bias voltage dependent, but within the [$-1$\,V;$+1$\,V] range it is at best slightly above the noise of most modern commercial STMs leveling at $\approx 2$\,pm~\cite{Brede2014}~\cite{Decker2014}. 
When turning to the resonant tunneling mode, both topographies recorded at $U=4.8$\,V and $5$\,V shown in Fig.~\ref{fig1}(c) and (d), respectively, exhibit an overall drop in corrugation between the top and the valley sites.
This can be seen most easily by the faint contrast in Fig.~\ref{fig1}(c), while a striking inversion of apparent corrugation is observed at $U=5$\,V.
Interestingly, a contrast between fcc and hcp sites at $U=5$\,V is evident. 
Fig.~\ref{fig1}(e) shows line profiles across the moiré structure measured for conventional and resonant STM, as marked in Fig.~\ref{fig1}(b) and (d), respectively.
The overall corrugation drops from $150$\,pm peak-to-peak (Pk-Pk) at $U=-0.5$\,V to just below $30$\,pm Pk-Pk for $U=5$\,V. 
At $U=5$\,V, the top sites are imaged as depressions and the ridges separating fcc and hcp turn into a plateau. 
Remarkably, the apparent height difference between fcc and hcp regions rises to $8$\,pm. 
In comparison to conventional STM, the combination of reduced overall roughness and large differences in apparent height in the resonant tunneling mode makes the contrast between the fcc and the hcp site distinctive.

\begin{figure}[tb]
\begin{center}
\includegraphics[width=0.5\columnwidth]{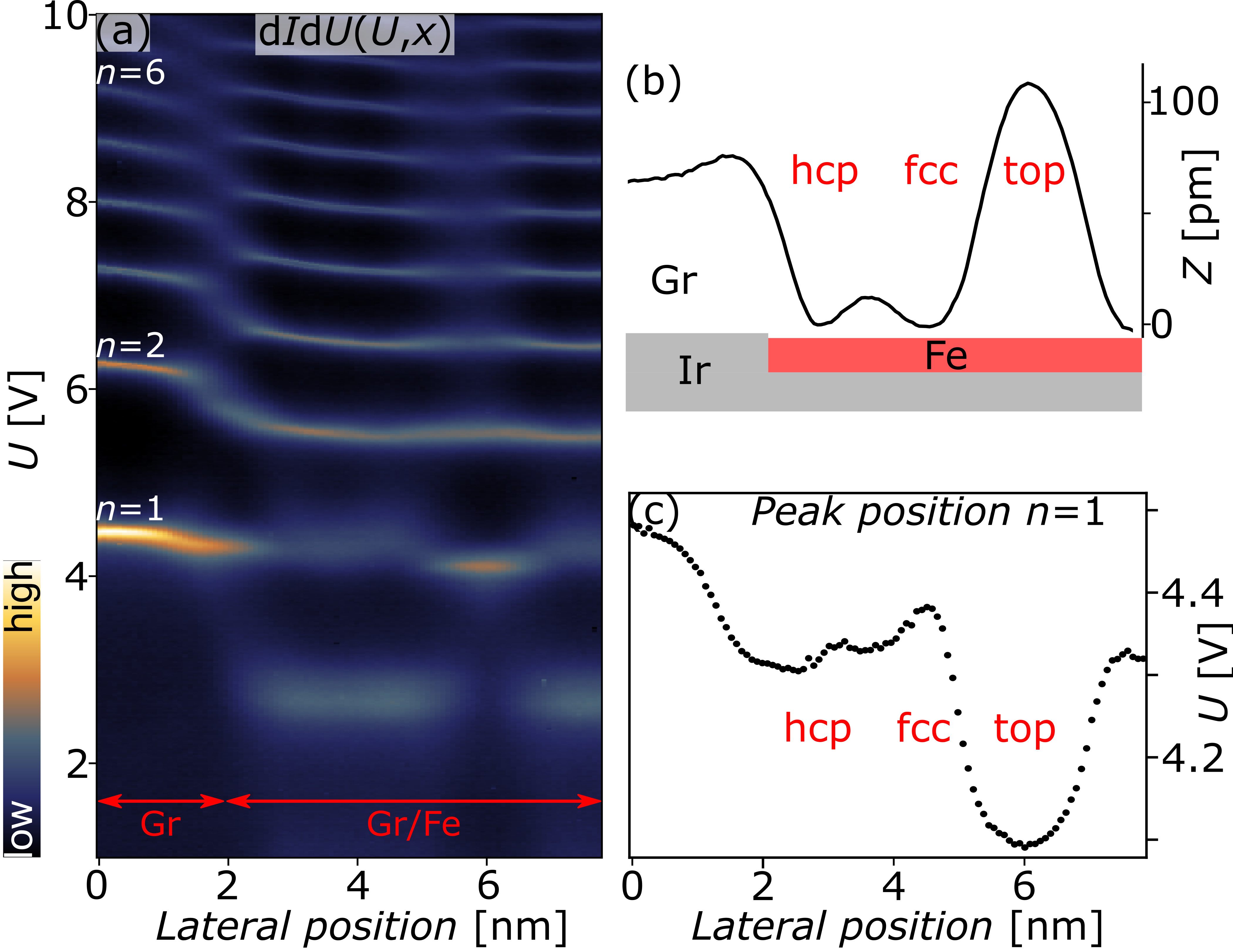}
\caption{\textbf{Evolution of IPSs across the lateral interface between Gr/Ir(111) and Gr/Fe/Ir(111).}
\textbf{(a)}~d$I$/d$U(U,x)$ recorded along the line marked in Fig. \ref{fig1}(b) by the purple arrow, revealing the IPS series on Gr and Gr/Fe regions (marked on waterfall plot) and their transition ($I=1$\,nA).
IPSs of order $n = 1,2,6$ are labeled on the Gr site. 
\textbf{(b)}~Scheme of the sample system shown in this figure together with an apparent height profile extracted from stabilization positions at $U=1$\,V of the spectra in (a).
The scheme shows Gr covering both Ir(111) and a monolayer of Fe on Ir(111) on the lower Ir terrace.
High symmetry positions hcp, fcc and on-top are marked.
\textbf{(c)}~Spatial evolution of the $n=1$ IPS peak position extracted by Lorentzian fits to data in (a).
 d$I$/d$U$ range for color bar: $[0; 1.2]$~\,arb.~u.
}
\label{fig2}
\end{center}
\end{figure}

To understand the origin of the pronounced stacking contrast observed in resonant tunneling mode, we turn to spatially-resolved investigations of IPSs above the Fe-intercalated Gr.
We start with a study of the evolution of the IPSs across the lateral interface between Gr/Ir(111) and Gr/Fe/Ir(111).
Figure~\ref{fig2}(a) shows a waterfall plot of the differential conductance~d$I$/d$U(U,x)$ taken along the purple arrow marked in the topography image shown in Fig.~\ref{fig1}(b). 
The scanned surface area comprises a buried Ir step resulting in a lateral interface between the upper Ir(111) terrace and an Fe monolayer on the lower terrace both being covered by Gr, as depicted in the schematic sketch in Fig.~\ref{fig2}(b). 
The corresponding topography profile along the arrow marked in Fig.~\ref{fig1}(b) is shown as well. 
On the pristine Gr/Ir(111) region, eight maxima can be found in the d$I$/d$U(U)$ curves of Fig.~\ref{fig2}(a) revealing the lowest eight IPSs.
When going from Gr/Ir(111) to the Fe intercalated region, all IPSs shift to lower energies, resulting in the appearance of the $9^{\rm th}$ IPS slightly below $10$\,V on Gr/Fe/Ir(111). 
Here, the smooth spatial transition between the two IPS series via the buried interface allows unambiguous assignment of each order~$n$. 
An additional maximum in the d$I$/d$U(U)$ at lower energies ($U=2.7$\,V) appears on the intercalated Gr region. 
Its intensity is strongest on the hcp and fcc areas, while almost diminishing on the on-top position. 
We attribute this feature to an interface state (IFS) originating from the hybridization of the Gr with the Fe. 
The occurrence of such an IFS is typical for strongly interacting Gr-metal systems~\cite{Borca2024, Armbrust2015, GarciaLekue2014, Gyamfi2012, Borca2010, Wang2010, Bose2010}. 
In Fig.~\ref{fig2}(c) the spatial evolution of the $n=1$ IPS has been analyzed in more detail, showing its peak position extracted from the d$I$/d$U(U)$ data shown in Fig.~\ref{fig2}(a). 
Following the first IPS peak position across the intercalated moiré unit cell from hcp, via fcc, on-top to another hcp site, its energy varies between all of them. 
Most significantly, the IPS for the on-top configuration exhibits the lowest energy, followed by the hcp, and fcc having the highest energy.
Apparently, the IPS's energy is sensitive to the local registry of Gr on the Fe/Ir(111).

\begin{figure}[tb]
\begin{center}
\includegraphics[width=0.5\columnwidth]{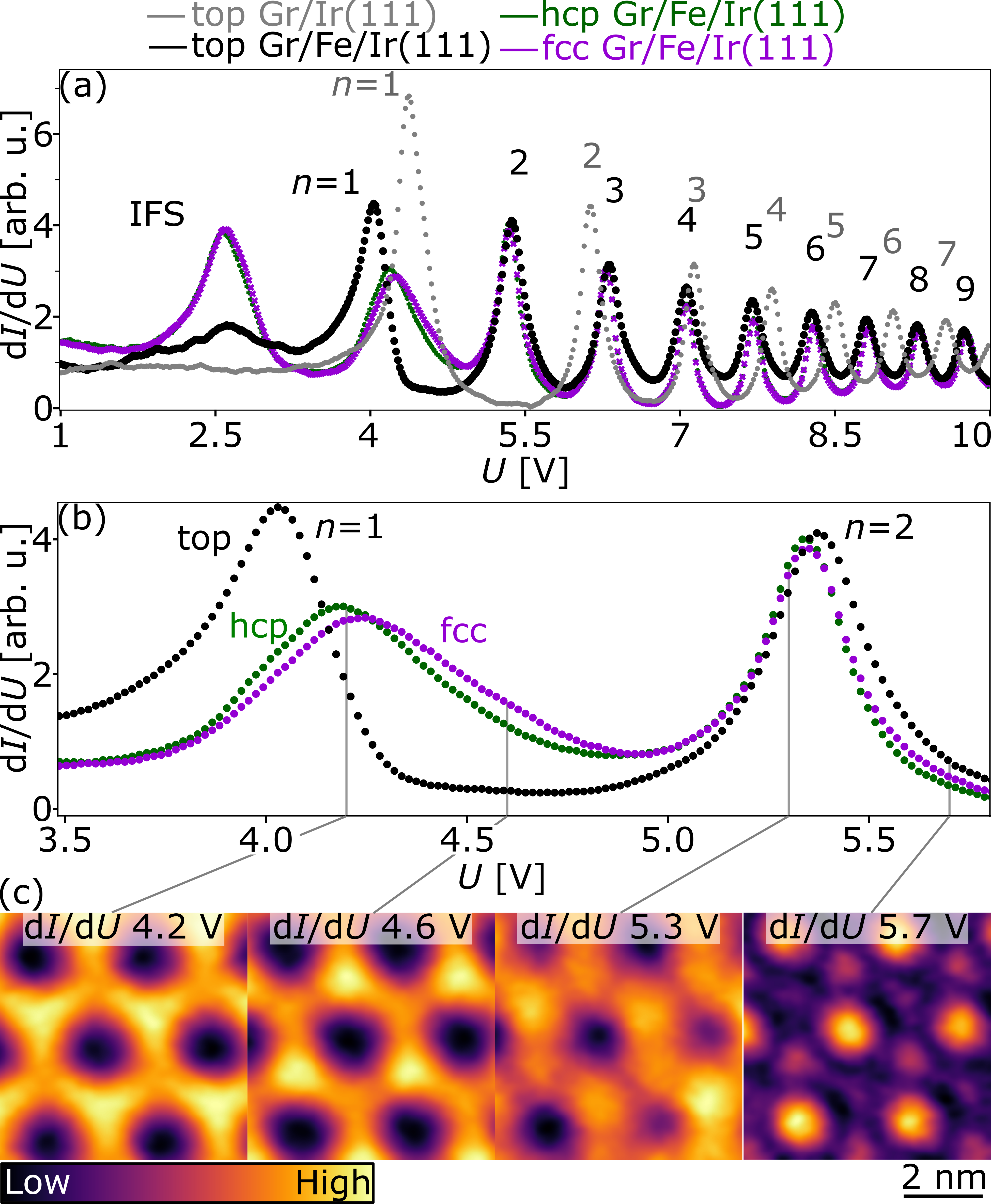}
\caption{\textbf{Sensing the local registry of Gr on Fe/Ir(111) by IPSs.}
\textbf{(a)}~d$I$/d$U(U)$ taken on high symmetry sites of the moiré unit cell on an extended Gr/Fe/Ir(111) region (hcp in green, fcc in violet, on-top in black) and on clean Gr/Ir(111) (gray). Each curve is an average of three consecutive scans.
\textbf{(b)}~A zoom in on $n=1$ and $n=2$ maxima of (a) highlighting the differences between the on-top, fcc and hcp positions. 
\textbf{(c)}~Series of d$I$/d$U(U)$ maps at indicated bias voltages exhibiting a contrast between all three high symmetry regions.
 ($I=1$\,nA, $B=1$\,T).
 d$I$/d$U$ range for color bars in corresponding maps from left: $[1.5; 3]$, $[0.1; 1.8]$, $[3.1; 4]$, $[0.1; 0.5]$~arb.~u..
}
\label{fig3}
\end{center}
\end{figure}

On close inspection of the entire IPS series it becomes evident, that even $6$\,nm apart from the buried step edge there is still an overall trend of shifting the peak positions to lower energies.
This is attributed to a combination of edge effects and stress in Gr close to the buried step edge.
Therefore, the difference in IPS energy between the different stackings displayed in Fig.~\ref{fig2}(c) might incorporate some additional energy shift due to the close-by interface.
To circumvent that we proceed on an extended intercalated region far away from any lateral interfaces.
In Fig.~\ref{fig3}(a) d$I$/d$U(U)$ curves ranging from $U=1$\,V to $10$\,V recorded on the three high symmetry sites of the moiré unit cell of Gr/Fe/Ir(111) are shown. 
For comparison, a reference spectrum for the on-top position of Gr/Ir(111), recorded with the same tip, is also shown.
The spectra reveal the lowest nine and seven IPSs of the series, respectively. 
In addition, on the intercalated area far away from any step edge the IFS at $U=2.7$\,V is also observed. 
As discussed for Fig.~\ref{fig2}(a), the moiré top site exhibits only its very week signature, while it is most pronounced on the hcp and fcc areas. 
This indicates a significant probability density of the IFS above the Gr layer at the latter sites, allowing for hybridization with the $n=1$ IPS~\cite{Borca2010}. 
Comparing the spectra on Gr/Fe/Ir(111) to the reference spectrum on Gr/Ir(111) the IPS series on each of the high symmetry points shifts to lower energy, as also observable in Fig.~\ref{fig2}(a).
IPSs are sensitive probes of alterations in the surface potential on electronically inhomogeneous surfaces, which manifest as spatial changes in the work function\cite{Jung1995, Borca2024, Niesner2014, Liu2021, Wang2010}.
Consequently, we attribute this shift to a work function reduction of Gr/Ir(111) upon Fe intercalaction (see section 1 of the supplemental material for further discussion). 
Using a simple model, as outlined therein, we find for the on-top position a reduction of the work function of Gr/Ir(111) by $\approx 0.9$\,eV upon intercalation of Fe, indicating a considerable charge transfer from Fe towards Gr.
Moreover, within the Gr/Fe/Ir(111) moiré unit cell we observe variations of the work function of up to $(62 \pm 4)$\,meV, the value laying well in between the lower limit given by $\approx 35$\,meV for the van der Waals–bound Gr/Ir(111)~\cite{Dedkov2014} and the upper limit of $(100-220)$\,meV for strongly bound systems like Gr/Ru(0001)~\cite{Borca2010, Brugger2009} or Gr/Rh(111)~\cite{Wang2010}.

Note, that the IPS $n=1$ on both valley sites is found at larger energies than on the top site, exhibiting an energy shift in the opposite direction to the local work function change. 
This splitting of the first IPS into a lower-energy constituent attributed to the top areas and a higher-energy constituent attributed to the valley sites has been observed before on various strongly interacting Gr-metal systems~\cite{Borca2024, Carnevali2023, Armbrust2015, Armbrust2012, Borca2010, Wang2010, Zhang2010}.
It is assigned to be a consequence of the high corrugation of these systems~\cite{Armbrust2015, Armbrust2012, Borca2010}: 
At small Gr-metal distances typical for the valley sites, the IPS wavefunctions almost resemble those of IPSs on a pure metal surface but with a higher degree of decoupling, since the Gr overlayer at smallest distance from the metal tends to push the IPS wavefunctions away from the metal surface, thereby reducing their binding energies~\cite{Armbrust2015, Borca2010}.
For Gr/Ru(0001), with Gr-metal spacing variations comparable to Gr/Fe/Ir(111), the dispersions observed by angle-resolved 2PPE are indicative of electrons moving almost freely above the spatially connected valley areas~\cite{Armbrust2012}.
In contrast, for the on-top sites with increased Gr-metal spacing, the metal surface and the Gr layer can act as a quantum well, with the first IPS partially trapped between the metal and the Gr.
Its wavefunction develops a local maximum in between the metal and Gr~\cite{Armbrust2015, Borca2010}, and is therefore more strongly coupled to the metal compared to the valley regions. 
This explains the larger binding energy, shorter lifetime, and higher effective mass of the $n=1$ IPS on the top sites compared to the valleys, as observed for Gr/Ru(0001)~\cite{Armbrust2012}, with the increased effective mass indicating additional lateral confinement at these sites~\cite{Armbrust2012,Borca2010}.
To conclude, based on previous results and our experimental observations, the $n=1$ IPS on the top site of the Gr/Fe/Ir(111) moiré structure is expected to attain properties of an interfacial state known to be closely related to the interlayer state of graphite and of other layered materials~\cite{Silkin2009,Blase1995,Cuong2014}. 
Note, that this corrugation effect is expected to dominate the $n=1$ IPS, while as $n$ increases, the states become increasingly delocalized perpendicular to the surface and thus progressively less affected by the Gr-Fe distance variation~\cite{Armbrust2015, Borca2010}.
In conclusion, the difference in the potential energy landscape for the IPS series on the high symmetry sites results in the observed different energetic positions of the IPSs within the moiré unit cell.
The resulting change of IPS peak position and height in the d$I$/d$U(U)$ spectroscopy curves, as clearly observable in the zoom-in shown in Fig.~\ref{fig3}(b), correlate with different d$I$/d$U(U)$ levels at a fixed bias $U$ on theses sites. 
These differences can be further spatially resolved by mapping the site-dependent differential conductance d$I$/d$U(U,x,y)$ at fixed bias $U$ in the resonant tunneling mode. 
In Fig.~\ref{fig3}(c) exemplary d$I$/d$U(U)$ maps recorded on Gr/Fe/Ir(111) are shown. 
The respective bias $U$ for the maps were selected to show the highest contrast between the high symmetry sites. 
At $U=4.2$\,V and $4.6$\,V the on-top sites appear dark, while the bright regions indicate the valleys with different intensities on fcc and hcp sites, respectively.
Note, that when going from $U=4.2$\,V to $4.6$\,V, the relative intensities on the valleys change: while at $U=4.2$\,V the brightest regions in the map indicate the hcp sites, at $U=4.6$\,V the brightest regions are the fcc sites.
As can be seen in Fig.~\ref{fig3}(b), this change in intensity is related to the small shift of the IPS peak position between hcp and fcc sites.
The observed differences in the local differential conductance are also responsible for the stacking contrast observed in the topography images shown in Fig.~\ref{fig1}, since in topography images recorded at a given bias voltage $U$ all tunneling channels are integrated from $E_{\rm F}$ up to an energy defined by e$U$.  
Notably, the stacking contrast in the d$I$/d$U$ maps and its inversion are also observable for the $n=2$ IPS as exemplary shown in Fig.~\ref{fig3}(c) for maps recorded at $U=5.3$\,V and $5.7$\,V. 

\begin{figure}[t!]
\begin{center}
\includegraphics[width=1\columnwidth]{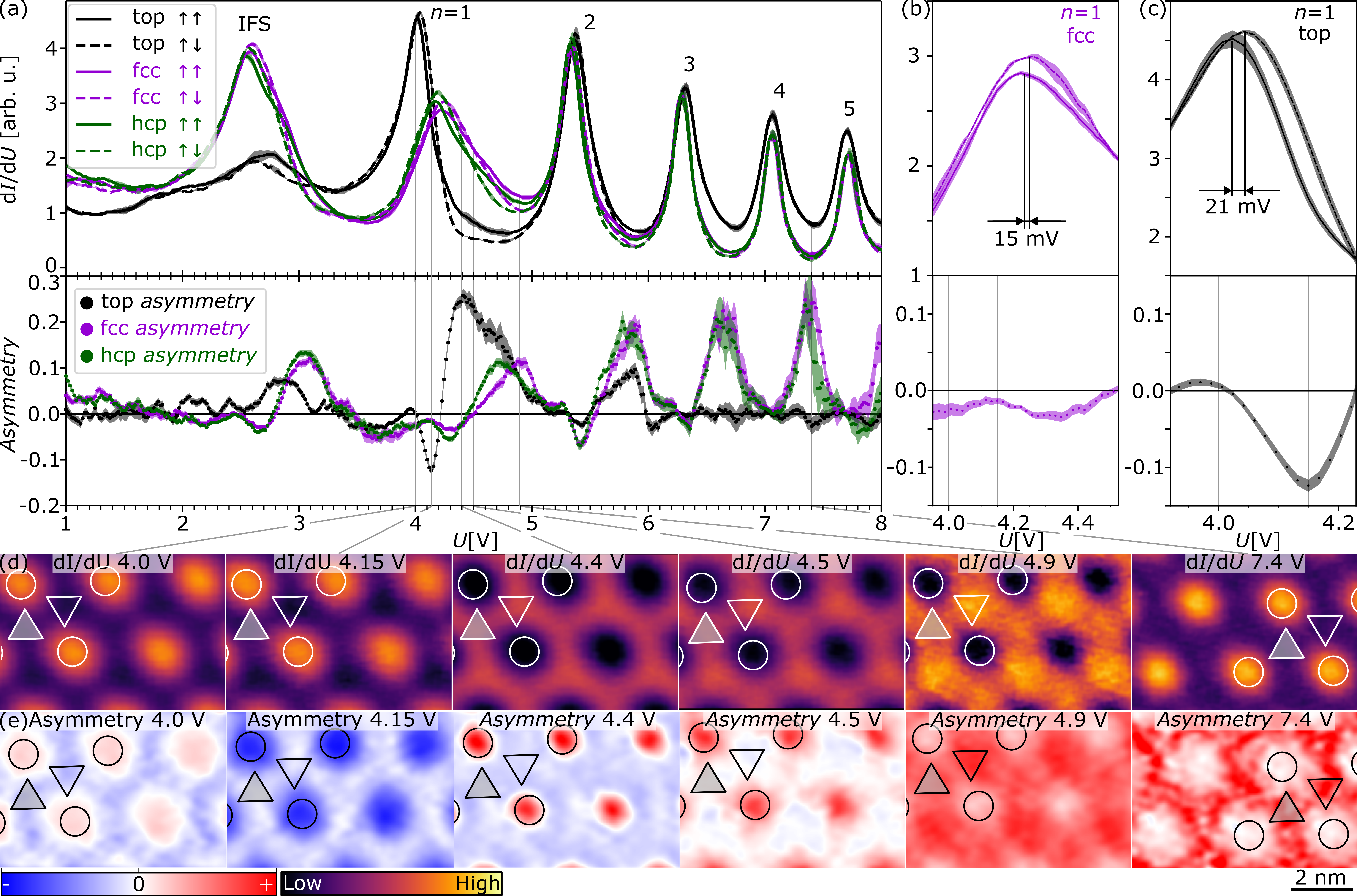}
\caption{\textbf{Spin contrast within the Gr/Fe/Ir(111) moiré unit cell mediated by IPSs.}
\textbf{(a)}~\textbf{Top:}~Spin-polarized tunneling spectra d$I$/d$U(U)$ recorded with an Fe-coated W tip for on-top, fcc and hcp sites in an external magnetic field of $B=\pm 1$\,T along the tip, respectively. 
The two configurations of magnetic moments of tip and sample measured are marked with solid and dashed lines. 
Six peaks in each curve are observed, related to the IFS and the lowest five IPSs.
Each curve is an average of three consecutive scans.
Shading areas show the standard deviation of the measured spectra.
\textbf{Bottom:}~Asymmetry plots computed from the above spectra. 
\textbf{(b)}~and \textbf{(c)}~A zoom in of (a) on the 1st IPS on the fcc and on-top position, respectively. 
Black vertical lines mark the position of fitted Lorentzian functions. The effective spin-split of the IPS is marked.
\textbf{(d)}~d$I$/d$U$ maps at indicated bias voltages recorded for the parallel alignment of tip and sample magnetic moments.
\textbf{(e)}~Corresponding asymmetry maps computed from d$I$/d$U$ maps registered in the parallel and antiparallel configurations of tip and sample magnetic moments.
The voltages at which the maps were recorded are marked in (a-c) by grey lines. 
The moiré unit cell is depicted on all maps. ($I=1$\,nA).
d$I$/d$U$ range for color bars in corresponding maps: from left $[1; 5.5], [2; 5.5], [0.8; 4], [0.8; 4.5], [0.6; 1.9], [-0.5; 1.2]$\,arb.~u..
Spin asymmetry range for color bars in corresponding maps: $[-0.15; 0.15]$ for the first two from the left, [-0.3; 0.3] else.
}
\label{fig5}
\end{center}
\end{figure}
In order to investigate the spin-polarization of the IPSs above the Gr-ferromagnet hybrid, we performed spin-resolved STS in the resonant tunneling mode on the fcc, hcp and on-top areas with parallel and antiparallel alignment of the tip and sample magnetization by means of external magnetic fields of $B=\pm 1$\,T, respectively. 
The respective spin-resolved d$I$/d$U(U)$ curves covering the energy range from $1$\,eV via the IFS up to the $5^{\rm th}$ IPS are shown in Fig.~\ref{fig5}(a).
Clearly, on each high symmetry site of the moiré unit cell the spin-dependent curves for the two magnetic configurations differ, indicating a spin-polarization of the IFS and the IPS series~\cite{Kubetzka2007,Schlenhoff2012} (see also section 4 of the supplemental material).
The calculated asymmetry of the spin-dependent d$I$/d$U(U)$ curves for each site is plotted in the lower panel of Fig.~\ref{fig5}(a). 
Defined by $Asymmetry(U)=($d$I$/d$U(U)_{\uparrow\uparrow}$ -d$I$/d$U(U)_{\uparrow\downarrow}$)/(d$I$/d$U(U)_{\uparrow\uparrow}$+d$I$/d$U(U)_{\uparrow\downarrow}$), with the arrow directions referring to the magnetization directions of the SP-STM probe tip and the sample, it indicates the fraction of the spin-polarized to the spin-averaged signal at a given bias $U$, and is thus a measure for the effective electron spin-polarization at the respective energy e$U$~\cite{Wortmann2001, Kubetzka2003, Wiesendanger2009} (see section 3 of the supplemental material for a detailed discussion). 
Obviously, the electron spin-polarization is strongly energy-dependent determined by the presence and energetic positions of the IFS and the IPSs (see section 4 of the supplemental material for a detailed discussion).
Moreover, a considerable spatial variation within the moiré unit cell is observed.

In Fig.~\ref{fig5}(b) a magnified view of the peaks of the spin-resolved d$I$/d$U(U)$ curves of Fig.~\ref{fig5}(a) corresponding to the $n=1$ IPS on the fcc position is shown. 
We find an effective spin-splitting of $(15\pm 3)$\,meV. 
For the on-top position, shown in Fig.~\ref{fig5}(c), the first IPS related peak is spin-split by even $(21\pm 3)$\,meV.
These values are comparable to the effective spin-splitting of IPSs observed by resonant SP-STS on thin-film metals on heavy-element substrates~\cite{Kubetzka2007,Schlenhoff2012}. 
Note, that on both sites, the d$I$/d$U(U)_{\uparrow\downarrow}$ peak is not only shifted to higher energies, but also exhibits a slightly different height and width, that may arise from a spin-dependent probability density and lifetime of the IPSs above this complex Gr-ferromagnet hybrid system~\cite{Passek1995, Nekovee1993}.
As shown in the lower panel of Fig.~\ref{fig5}(c), for the top site this results in a small local maximum of the asymmetry curve at $U=3.97$\,V, corresponding to the rising flanks of the spin-split d$I$/d$U(U)$ curves, and a pronounced minimum of $-0.12$ at $U=4.15$\,V, corresponding to the falling flanks of the spin-split peaks.
(For a detailed discussion of the effect of the asymmetric peak shapes on the resulting spin asymmetry see section 3 of the supplemental material.)
Hence, the spin-split IPS on the top site causes an effective spin-polarization of up to $12\,\%$ in this narrow energy range around its peak maximum. 
For the fcc site, as observable in Fig.~\ref{fig5}(b), the entire d$I$/d$U(U)_{\uparrow\downarrow}$ curve runs above the d$I$/d$U(U)_{\uparrow\uparrow}$ curve around the respective spin-split IPS peak maxima, causing only negative asymmetries in this narrow energy range. 
As deduced from the minimum in the asymmetry curve at $U=4.31$\,V, shown in the lower panel, an effective spin-polarization of $<3\,\%$ is found for the $n=1$ IPS in the respective energy range.   
Analogous to the fcc site, on the hcp site we find an effective spin-polarization of $<6\,\%$ for the $n=1$ IPS at $U=4.31$\,V close to its spin-split peak maxima (see Fig.~\ref{fig5}(a)). 
Consequently, close to its respective maxima in the spin-split d$I$/d$U(U)$ curves, the $n=1$ IPS exhibits a factor of $4$ ($2$) larger spin-polarization on the moiré top site then on the fcc (hcp) site.  

In order to map the effective IPSs' spin-polarization across the moiré unit cell, d$I$/d$U$ maps on the very same area have been recorded in the resonant tunneling mode once for parallel and once for antiparallel alignment of tip and sample magnetization and the corresponding asymmetry maps were deduced. 
Fig.~\ref{fig5}(d) shows exemplary d$I$/d$U$ maps recorded at different bias voltages, as marked in Fig.~\ref{fig5}(a-c).
The corresponding spin asymmetry maps are shown in Fig.~\ref{fig5}(e). 
In the first map taken at $U=4.0$\,V, we observe a small positive spin asymmetry on the top sites, and a small negative asymmetry in the valleys, in agreement with the point spectroscopy data shown in Fig.~\ref{fig5}(a-c). 
The second map has been recorded at $U=4.15$\,V, which corresponds to the minimum in the asymmetry curve for the top site, best observable in Fig.~\ref{fig5}(b).
Here, an intense negative spin asymmetry is mapped on the top sites, while the surrounding valleys exhibit an only weak spin-polarization of same polarity, consistent with the point spectroscopy curves of Fig.~\ref{fig5}(a-c). 

Interestingly, when going in the spin asymmetry curve for the top site from the minimum at $U=4.15$\,V slightly higher in energy to $U=4.4$\,V (see Fig.~\ref{fig5}(a)), we observe another pronounced maximum, located at the spin-dependent long tails of the spin-split $n=1$ IPS peaks in d$I$/d$U(U)$.
It arises from the slightly different shapes of the d$I$/d$U(U)_{\uparrow\uparrow}$ and d$I$/d$U(U)_{\uparrow\downarrow}$ curves for the two magnetic configurations, best observable at the asymmetric falling flanks and long tails of the respective spin-split peaks (see sections 3 and 4 of the supplemental material for a detailed discussion). 
For the same reason, we observe a pronounced maximum in the spin-asymmetry at $U=4.9$\,V ($U=4.7$\,V) for the fcc (hcp) site, located at the spin-dependent long tails of the respective $n=1$ IPS peaks. 
At those respective maxima, we find a large asymmetry of $26$\,$\%$ induced by the IPS $n=1$ on the moiré top site, and a smaller asymmetry of $11\,\%$ for both valley sites. 
Consequently, the same trend as close to the IPS-induced spectroscopic peak maxima is observed:  
The spin-polarization on the moiré top site is found to be larger then for the valleys. 

For comparison, our conventional SP-STS measurements of states around $E_{\rm F}$ in a bias range of [-1\,V; +1\,V] on the same sample with the same probe tip reveals an asymmetry of up to $42$\,$\%$ on both valley sites (see section 2 of the supplemental material), consistent with previous studies~\cite{Decker2014,Brede2014} and attributed to the strong Gr-Fe hybridization on these sites. 
The states on the moiré top site, on the other hand, exhibit an only weak spin-polarization with an asymmetry of $\leq10$\,$\%$ reflecting the physisorbed character of Gr on Fe/Ir(111) accompanied by an only weak spin transfer from the Fe to the Gr on this position (see section 2 of the supplemental material), as previously studied~\cite{Decker2014, Brede2014}.
Consequently, the states around $E_{\rm F}$ exhibit a factor of 4 smaller spin-polarization on the moiré top sites compared to the valleys. 

Moreover, comparing the spin asymmetry of up to $42$\,$\%$ on both valley sites (see section 2 of the supplemental material) with the $11\,\%$ induced by the $n=1$ IPS, we observe a reduction of the spin-dependent signal by up to $74\,\%$ when changing from the direct to the resonant tunneling mode.
Such a strong decrease in the magnetic signal yet appeared to be typical when changing to spin-polarized resonant tunneling through IPSs, as this trend has been observed on all the previously studied systems~\cite{Schlenhoff2020, Schlenhoff2019, Schlenhoff2012, Kubetzka2007}. 
However, with a maximum asymmetry of $26$\,$\%$ induced by the $n=1$ IPS peak on the on-top sites of this complex Gr/Fe/Ir system, we observe for the first time a higher effective electron spin-polarization compared to the states around $E_{\rm F}$  (exhibiting an asymmetry of $\leq10\,\%$), and accordingly an  enhancement of the spin-dependent signal when turning to the resonant tunneling mode. 

The observation of a spin-polarization of the $n=1$ IPS following a moiré site-specific trend opposite to the one of the states around $E_{\rm F}$ and of the enhancement of the spin signal on the top sites when turning to resonant tunneling 
is surprising only at first glance, considering the interfacial character of the $n=1$ IPS on the top site (discussed above) and the different nature of the states around $E_{\rm F}$ involved in direct tunneling:
Because of the physisorbed character of Gr in the local on-top registry, the largest spin-density is concentrated in the Fe layer below the Gr, as known from previous theoretical and experimental work~\cite{Brede2014}.
Naturally, conventional SP-STS is most sensitive to the evanescent tail of spin-polarized states localized in the very top-most layer, thereby sensing the small (large) Gr spin-polarization on the top-site (valleys)~\cite{Brede2014, Decker2014}.

In contrast, resonant SP-STS detects the IPSs in front of the surface, which, due to their interfacial character and ability to penetrate the sub-surface region below the Gr, may reflect not only charge accumulation but also spin accumulation at buried interfaces, as demonstrated here.
Consequently, our observations indicate a significant penetration of the first IPS into the Gr/Fe/Ir(111) for the moiré top sites and a spin-dependent scattering at the buried Fe layer. 

When considering the IPS $n=2$ related extrema in the spin asymmetry, shown in Fig.~\ref{fig5}(a), the site-specific trend of the spin-polarization appears to reverse, following now the same trend as observed for the states around $E_{\rm F}$: 
Close to the $n=2$ IPS peak maxima at $U=5.45$\,V, the spin-polarization on the valleys is larger ($7\,\%$) then on the top ($2\,\%$).  
Accordingly, at their spin-dependent long tails at $U=5.9$\,V ($U=5.8$\,V), the fcc (hcp) site exhibit an asymmetry of $19\,\%$ ($20\,\%$), but the top site only $10$\,$\%$. 
This is reasonable, since only the first IPS on the top site exhibits predominantly an interfacial character~\cite{}, as discussed above.
With increasing $n$, the IPSs expand further and further into vacuum, thereby sensing progressively less the highly polarized interface below Gr and in effect their spin-polarization drops off.   
However, since on the valleys the IPSs sense the magnetism induced in Gr rather than in the interface below, the drop off is smaller and the spin-polarization induced by the $n=2$ IPS reflects the larger Gr spin-polarization on these sites. 
The fact, that on the valleys the $n=2$ IPS exhibits an even larger spin-polarization then the $n=1$, we attribute to a strong hybridization of the latter with the IFS, as discussed above.  
When comparing the respective minima and maxima in the asymmetry curves in Fig.~\ref{fig5}(a) for the IFS and the $n=1$ IPS on the valley sites, both states exhibit a spin-polarization of same magnitude, proving their strong hybridization. 

For the $n>2$ IPSs on each moiré site, a spin-splitting of the peaks in d$I$/d$U(U)$ cannot be resolved anymore, indicating a decrease of the spin-splitting with increasing order, as generally expected from the decreasing overlap of the IPS wavefunctions with a spin-polarized surface~\cite{Donath2007}.
However, especially on the valleys, the d$I$/d$U(U)$ curves still exhibit a spin-dependence in terms of slightly different shapes (e.g. height, width, tail), resulting in pronounced spin asymmetry maxima in the regions where the long tails of the $(n-1)$ IPS peak overlaps with the rising flanks of the $n$ IPS peak (see section 3 of the supplemental material for a detailed discussion). 
In the measured bias range, this leads to a spin asymmetry of up to  $24\,\%$  at close to $7.4$\,V, while the signal to noise ratio gets worse as well at these high voltages.
Still, this high energy effective spin-polarization can be mapped, as demonstrated by the exemplary asymmetry map at $U=7.4$\,V shown in Fig.~\ref{fig5}(e).
As discussed before, the effective spin-polarization on the on-top positions is negligible while it is positive for fcc and hcp in accordance with the asymmetry curves shown in Fig.~\ref{fig5}(a). 

Finally we want to highlight, that while the fcc and hcp areas are indistinguishable in the low bias spin asymmetry deduced from direct tunneling data~\cite{Brede2014, Decker2014}, in resonant SP-STS we observe fine differences in the effective spin-polarization between those different valley sites. 
In Fig.~\ref{fig5}(a), a considerable difference in the spin asymmetry for the respective sites is found between the $n=1$ and $n=2$ IPS related d$I$/d$U(U)$ peaks.  
Here hcp exhibits a higher spin-polarization at a bias of $U=4.7$\,V, while the fcc spin-polarization dominates at $U=4.9$\,V. 
Evidently, the spin contrast between the high-symmetry sites in resonant SP-STS at a fixed bias voltage arise from the site-dependent slightly different energetic positions of the spin-polarized IPSs. 
In the spin asymmetry map at $U=4.4$\,V, shown in Fig.~\ref{fig5}(d), we observe the highest electron spin-polarization on the on-top position and a sign inversion between the on-top and the valley sites. 
Such a sign inversion has been previously observed also for states around $E_{\rm F}$~\cite{Decker2014,Decker2013}. 
However, the asymmetry map at $U=4.5$\,V exhibits the most pronounced spin contrast between the fcc and hcp positions. 
Here, the asymmetry for hcp has a positive sign while for fcc it is negative. 
At the same time the on-top sites exhibit a very strong positive spin-polarization. 
At $U=4.9$\,V all three regions exhibit a positive electron spin-polarization in the asymmetry map. 
However, as also evident from the spectroscopy shown in Fig.~\ref{fig5}(a), the spin asymmetry for on-top and hcp site is lower than for the fcc, while for the latter it is the highest in the energy range between $n=1$ and $n=2$.
To conclude, our experiments demonstrate a moiré site- and energy-dependent effective spin-polarization of the IPSs on the Gr-ferromagnet hybrid that can be mapped across the entire moiré unit cell. 

\section{Conclusions}
We were able to show a contrast in high-bias resonant tunneling images and spectroscopy between fcc and hcp stacking sites of the moiré unit cell in a Gr-based 2D hybrid system, unachievable in conventional low bias STM. 
The contrast originates from the spatial variation of the IPSs, being sensitive to the local surface work function variation within the moiré unit cell, determined to be in the order of $(62 \pm 4)$\,meV, as well as to the strong Gr-Fe distance variation.
The spin-polarized measurements reveal a moiré site- and energy-dependent effective IPSs' spin-polarization, providing a spin contrast between hcp and fcc sites unobtained in conventional SP-STM/STS.
Moreover, the first two IPSs exhibit a huge spin-polarization on the top-sites, where the physisorbed Gr is known to exhibit only a weak spin-polarization.
This finding implies a significant probability density of these IPSs below the Gr, sensing the spin-density in the buried Fe layer.
Consequently, our work demonstrates that the lowest order IPS can be used to locally sense the spin density at magnetic interfaces buried by a non-magnetic passivation layer. 
In light of the rising field of twistronics where the interface of stacked 2D systems and the interplay of charge, spin and moiré superlattice structure is decisive for emerging phenomena, atomic-scale investigations of IPSs may provide further insight into these buried processes.

\section{Methods}
All experiments were performed in an UHV system equipped with a low temperature SP-STM and two preparation chambers for substrate cleaning, Gr CVD growth, and metal deposition~\cite{Wittneven1997}.
The Ir(111) single crystal was cleaned by repeated cycles of Ar$^{+}$ sputtering ($800$\,V, $5$E$-6$\,mbar), annealing at temperatures ranging from $T=900$\,K to $1500$\,K in the presence of O$_{2}$ and a flash annealing at $T\approx 1500$\,K. 
The Gr layer was grown in situ on Ir(111) by thermal decomposition of ethylene molecules following the procedure described in ref.\cite{NDiaye2008}. 
Less than monolayer Fe intercalation was achieved in situ following the procedure of ref.\cite{Bazarnik2015}. 
After preparation samples were transferred in vacuo to the SP-STM setup and cooled down to the measurement temperature of $T=6.5$\,K. 
A lock-in detection technique was used to obtain d$I$/d$U$ maps and point spectroscopy data adding a small AC modulation $U_{\rm {Pk-Pk}}=50$\,mV to the bias voltage~$U$ at $f=1111.1$\,Hz. 
The d$I$/d$U$ maps were recorded simultaneously with the STM topography in the constant-current mode.
High bias differential conductance spectra have been recorded while maintaining a constant current setpoint during the bias voltage sweep up to $U=10$\,V~\cite{Binnig1985, Becker1985, Schlenhoff2020}. 
IPS position in point spectra was extracted by fitting Lorentzian function to each peak. 
The work function was determined by fitting to $n=<3, 6>$. 
Electrochemically etched tungsten tips cleaned by standard in vacuo flash annealing at $T\approx 2500$\,K and coated with $\sim50$\,atomic layers of Fe were used as probes for our SP-STM studies.
These Fe/W tips are soft ferromagnets and can be easily aligned with an external magnetic field. 
The sample is a hard magnet with an out-of-plane easy axis and cohersive field around $6$~T~\cite{Sierda2019, Decker2014, Brede2014}. 
Therefore, applying fields along the axis of the tip with strengths up to $2$~T we can easily control the alignment of the magnetic moment of the tip while not changing the alignment of magnetic moments on the sample.
The spin asymmetry is defined here as Asymmetry=(d$I$/d$U$↑↑-d$I$/d$U$↑↓)/(d$I$/d$U$↑↑+d$I$/d$U$↑↓) where the arrow directions refer to the magnetization direction of the SP-STM probe tip and the sample.
The error bar shown in Fig.~\ref{fig5}(a-c) is determined from a control experiment in which the tip magnetization has been flipped back to its initial state and an asymmetry of both the first d$I$/d$U$↑↑ and last  d$I$/d$U$↑↑ measurements was computed.
All data has been processed using self-written python code and Gwyddion\cite{Necas2012} software.

\section{Authors contribution}
MB and AS conceived the experiment, MB performed the SP-STM experiments, MB and AS analyzed the data, discussed the results and wrote the manuscript.

\begin{acknowledgement}
The authors would like to thank R.~Wiesendanger for support and provision of laboratories and infrastructure. 
AS would like to thank M. Donath for valuable discussions.
\end{acknowledgement}

\providecommand{\latin}[1]{#1}
\makeatletter
\providecommand{\doi}
  {\begingroup\let\do\@makeother\dospecials
  \catcode`\{=1 \catcode`\}=2 \doi@aux}
\providecommand{\doi@aux}[1]{\endgroup\texttt{#1}}
\makeatother
\providecommand*\mcitethebibliography{\thebibliography}
\csname @ifundefined\endcsname{endmcitethebibliography}
  {\let\endmcitethebibliography\endthebibliography}{}

\end{document}


\maketitle
\section{Determination of local work functions for Gr/Ir(111) and Gr/Fe/Ir(111)}
Image-potential states (IPSs) form a Rydberg-like series that is bound to and converges on the vacuum level of the surface. 
Hence, a change of the sample work function leads to a change in the IPS binding energies~\cite{Echenique1978}.
In scanning tunnelig microscopy (STM), due to the electric field applied to the surface, the IPSs experience a Stark shift that continuously shifts and expands the series~\cite{Crampin2005}.
Still, the IPSs are determined by the potential in front of the surface that can be approximated by a superposition of a linear electrostatic potential and the image-potential-induced surface potential~\cite{Binnig1985, Becker1985}.
Thus, they serve as probes for alterations in the surface potential, which manifest as changes in the work function. 
This relation has been proven to be useful to map surface work function changes on various sample systems by scanning tunneling spectroscopy (STS)~\cite{Jung1995, Pivetta2005, Ploigt2007, Schulz2014, Borca2020, Aladyshkin2020, Liu2021}, and in particular to probe the Gr-substrate interaction and resulting (local) work function changes~\cite{Niesner2014, Borca2024}. 
In the presence of an underlying substrate the work function of Gr is predominantly determined by charge transfer, which is driven by the occuring work function difference between Gr and the substrate~\cite{Niesner2014}. 
As a consequence, the analysis of IPSs opens the path to study interfacial charge transfer at the nanoscale~\cite{Liu2021}.
\begin{figure}[!htb]
	\centering
		\includegraphics[width=\columnwidth]{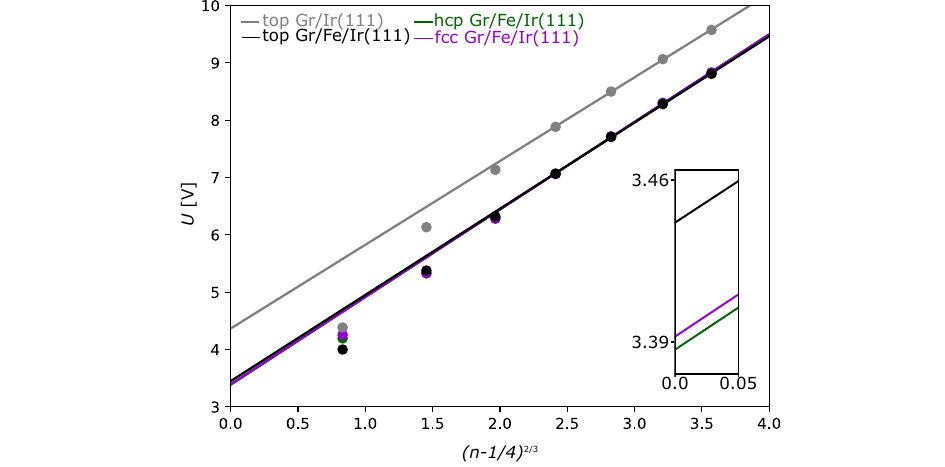}
		\caption{\textbf{Work function determination.} 
	IPS energy $U_{n}$ as a function of $(n-1/4)^{2/3}$ and line fits used to determine the local work function.
	(Peak positions $U_{n}$ are deduced from the d$I$/d$U(U)$ spectra taken on high symmetry sites of the moiré unit cell on Gr/Fe/Ir(111) (hcp in green, fcc in violet, on-top in black) and on clean Gr/Ir(111) (gray), presented in Fig.~3a of the main manuscript.)
\textbf{Inset:}~A zoom on the y-intercepts of the lines, yielding the work function values for the intercalated region.
		}
	\label{work_function}
\end{figure}

Comparing the d$I$/d$U(U)$ spectra on Gr/Fe/Ir(111) to the reference spectrum on Gr/Ir(111), shown in Fig.~3(a) in the main text, the IPS series on each of the high symmetry points shifts to lower energy.
We attribute this shift to a work function reduction of Gr/Ir(111) upon Fe intercalation. 
This can be further analysed using a simple model. 
Assuming an homogeneous electric field in front of the surface for the IPSs of order $n=[3;7]$ observed in constant current spectroscopy, 
their energetic positions $U_{n}$ can be approximated by a linear function of $(n-0.25)^{2/3}$, with the y-intercept yielding a rough value for the surface work function~\cite{Schlenhoff2022}.
In Fig.~\ref{work_function} the respective plots and fit curves for the three different high symmetry sites on Gr/Fe/Ir(111) as well as for Gr/Ir(111) are shown.
For the on-top position of the pristine Gr/Ir(111) we find $\phi_{\rm Gr/Ir}^{\rm top}=(4.364 \pm 0.003)$\,eV, which is slightly lower then the $(4.65\pm0.1)$\,eV obtained by two-photon photoemission~\cite{Niesner2012} and the $4.422$\,eV obtained by DFT calculations~\cite{Dedkov2014}. 
However, when going to the respective on-top position of the Fe-intercalated region a considerable lowering of the work function by $\approx 0.9$\,eV is evident in the data shown in Fig.~\ref{work_function}, yielding $\phi_{\rm Gr/Fe/Ir}^{\rm top}=(3.447 \pm 0.003)$\,eV and indicating a considerable charge transfer from Fe towards Gr. 
While this has not been studied before, upon intercalation of Cobalt resulting in a similar system a comparable decrease of the work function to about $3.3$\,eV has been reported~\cite{Calloni2020}.

Furthermore, within the Gr/Fe/Ir(111) moiré unit cell we observe variations of the work function of up to $(62 \pm 4)$\,meV.
Compared to the value on the top site, the work function for the fcc site is lower ($\phi_{\rm Gr/Fe/Ir}^{\rm fcc}=(3.391 \pm 0.003)$\,eV) and the lowest for the hcp site ($\phi_{\rm Gr/Fe/Ir}^{\rm hcp}=(3.385 \pm 0.003)$\,eV), in agreement with a strong hybridization of Gr with Fe/Ir(111) on these valley sites and the smallest Gr-Fe-distance on the hcp site~\cite{Decker2014,Brede2014}.
The observed work function variation within the moiré unit cell lies well in between the lower limit given by about $35$\,meV for the van der Waals–bound Gr/Ir(111)~\cite{Dedkov2014} and the upper limit of $(100-220)$\,meV for strongly covalently bound systems like Gr/Ru(0001)~\cite{Borca2010, Brugger2009} or Gr/Rh(111)~\cite{Wang2010}.

\section{Reference SP-STS study of electronic states around the Fermi level}
We recorded spin-polarized constant height tunneling spectra d$I$/d$U(U)$ between $U=-1$\,V and $+1$\,V taken on the exact same sites, respectively, as for the spin-polarized high-bias resonant tunneling spectra d$I$/d$U(U)$ presented in Fig.~4(a) in the main text. 
In order to exclude any tip-related artifacts, we used the exact same tip. 
The spectra were registered in parallel and antiparallel configurations of magnetic moments alignment between the tip and the sample and a corresponding spin asymmetry was computed.
A broad state with a local maximum in d$I$/d$U(U)$ at $U=-0.175$\,V is present in parallel magnetic moment configuration of tip and sample on all three moiré sites. This state is highly spin-polarized for hcp and fcc positions and therefore diminishes in the antiparallel configuration. 
In line with previously reported data~\cite{Decker2014, Sierda2019}, the highest spin polarization is visible between $U=-0.2$\,V and $-0.3$\,V. Both of the valley sites reach a 42$\%$ spin asymmetry and within the noise level are indistinguishable, while the on-top site has a circa 10$\%$ spin polarization. Note the change of sign between the valley sites and the on-top position at $U=-0.5$\,V.
\begin{figure}[!htb]
	\centering
		\includegraphics[width=0.8\columnwidth]{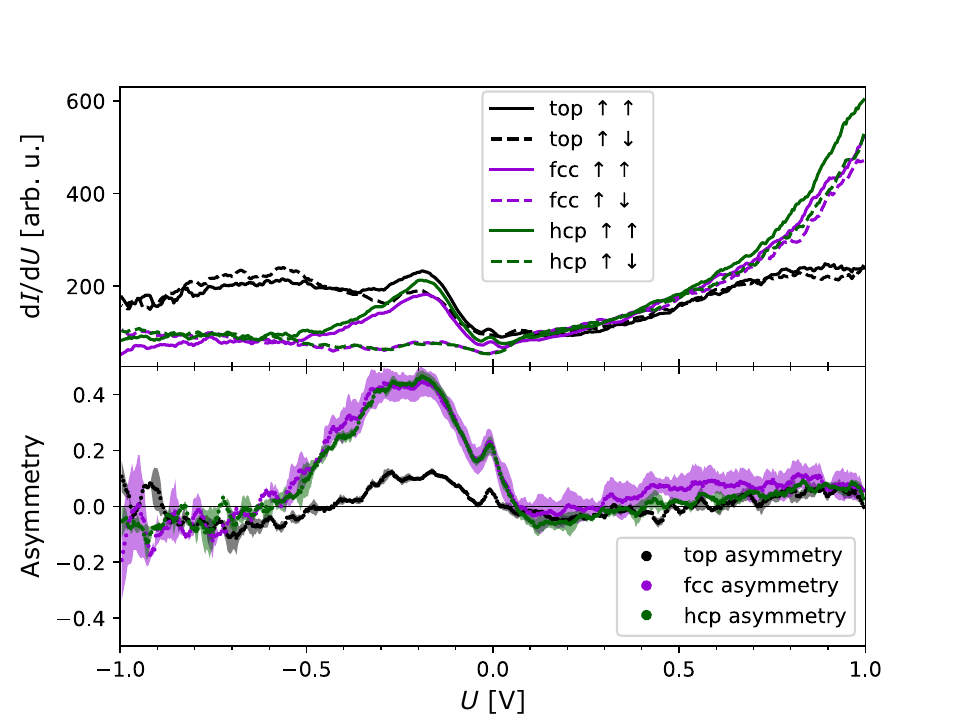}
		\caption{\textbf{Spin polarization near Femi energy.} 
		\textbf{Top:}~Spin-polarized constant height tunneling spectra d$I$/d$U(U)$ registered with an Fe-coated W tip for on-top, fcc and hcp positions in external magnetic fields of $\pm 1$\,T along the tip, respectively. Two configurations of magnetic moments of sample and tip measured are marked with solid and dashed lines. 
		\textbf{Bottom}:~Asymmetry plot computed from the above spectra. The error bar comes from a control experiment in which the tip magnetization was flipped to its original state and an asymmetry of both the first and last measurement was computed. Stabilization parameters: $U=1$\,V, $I_{\rm t}=4.5$\,nA.
}
	\label{sp_low_bias}
\end{figure}
\newpage
\section{Validation of spin asymmetry in spin-resolved IPS spectroscopy}\label{assy_section}
\subsection{Introduction of the concept}\label{asy_intro}
Spin asymmetry defined as the ratio of the difference of differential conductance of parallel and antiparallel configurations of the tip and sample magnetization to the sum of both signals was introduced to SP-STM data analysis by Kubetzka \textit{et al.} in their comprehensive work~\cite{Kubetzka2003}.
Normalizing the difference by the spin averaged signal under the assumption of a constant spin polarization of the tip allows for determination of the spin polarization of the local density of states (DOS) of the sample.
In standard scanning tunneling spectroscopy performed at constant height the prerequisite for the measurement is that the stabilization of the tip is performed at tunneling conditions at which the integrated spin polarized current is vanishing~\cite{Kubetzka2003, Wiesendanger2009}.
If this condition is not met, the starting position of the tip above the surface will depend on the magnetic moment orientation and the obtained difference in differential conductance and thus the spin asymmetry are flawed.
Note, that the tunneling current that determines the apparent topography is integrating all of the tunneling channels from the Fermi level to an energy e$U$ given by the applied bias $U$.
If the difference between integrated spin polarized currents for parallel and antiparallel configurations of the tip and sample magnetization (and the resulting topographic height difference) are non-vanishing, the tip stabilization height for the spectroscopy will differ which might lead to tunneling from or into different states of the sample due to their possibly different decay lengths into vacuum.

To obtain a spectrum of a series of Stark-shifted IPSs a constant height approach is no longer feasible, and constant current mode spectroscopy is required.
This is rooted in several physical and practical reasons.
Most significantly, in constant height mode, when ramping the bias voltage from values below to values exceeding the tip work function, field emission sets in, the transmission increases considerably, and the field-emission-induced contribution to the current strongly rises with increasing bias leading to an overload of the preamplifier. 
Consequently, while constant height spectroscopy of an individual IPS can still be realized, accessing the large energy range covering several members of a Stark-shifted IPS series within one measurement is only feasible in constant current mode. 
Here, since the feedback is regulating on a constant current, the tip is always retracting when ramping the bias up. 
Note, that this also assures an approximately constant electric field at the tip apex, and concomitantly a comparable electrostatic potential for the (low and intermediate) members of the Stark-shifted IPS series~\cite{Schlenhoff2022}.

Once, the applied bias allows for resonant tunneling into the first IPS, an additional transmission channel via tunneling into the IPS opens, thereby resulting in an tremendous increase in the overall current that is compensated by an appropriate, significant retraction of the tip manifesting as a step in the $z(U)$ characteristic (see Fig. \ref{combination_top}).
For the moiré top position of Gr/Fe/Ir(111), for example, when the applied bias equals the energy of the first Stark-shifted IPS, the tip retracts so far (by approximately $200$\,pm when passing $n=1$) that tunneling directly to the unoccupied density of states of the sample is highly unlikely with probability close to zero. 
This is easy to understand when considering a constant height spectroscopy over an IPS stabilizing the tip at a bias above the energy of the state.
When the sweep goes over the state to lower bias, the current drops to zero~\cite{SanderOtte2022}, indicating that the only contribution to the stabilization current at these distances comes from the IPS.
Consequently, we can rule out any contribution of direct electron tunneling from the tip to the surface at energies higher than the first IPS, and hence to our observed spin-asymmetry in between the IPSs.

When investigating the spin polarization of IPSs it is also important to start at a bias with vanishing integrated spin current.
Only if the starting difference in topography is zero, we can follow the evolution of the spin-dependent tip displacement $z(U)$ and attribute its difference $\Delta z$ for parallel and antiparallel magnetic configuration of tip and sample to the difference in the integrated spin currents.
In our case this condition is fulfilled at $U=1$\,V, because a small negative spin asymmetry close to the Fermi level is countered by a small positive asymmetry in the range of $[0.5; 1]$\,V (see the spectroscopy in Fig.~\ref{sp_low_bias}).
This allows us to set the starting tip-to-sample distance $z_0$ to $0$ and measure the change in $z$ as a function of bias for both tip-sample magnetic configurations.

\subsection{Symmetric spin-split peaks}\label{symmetric}
\begin{figure}[!htb]
	\centering
		\includegraphics[width=0.6\columnwidth]{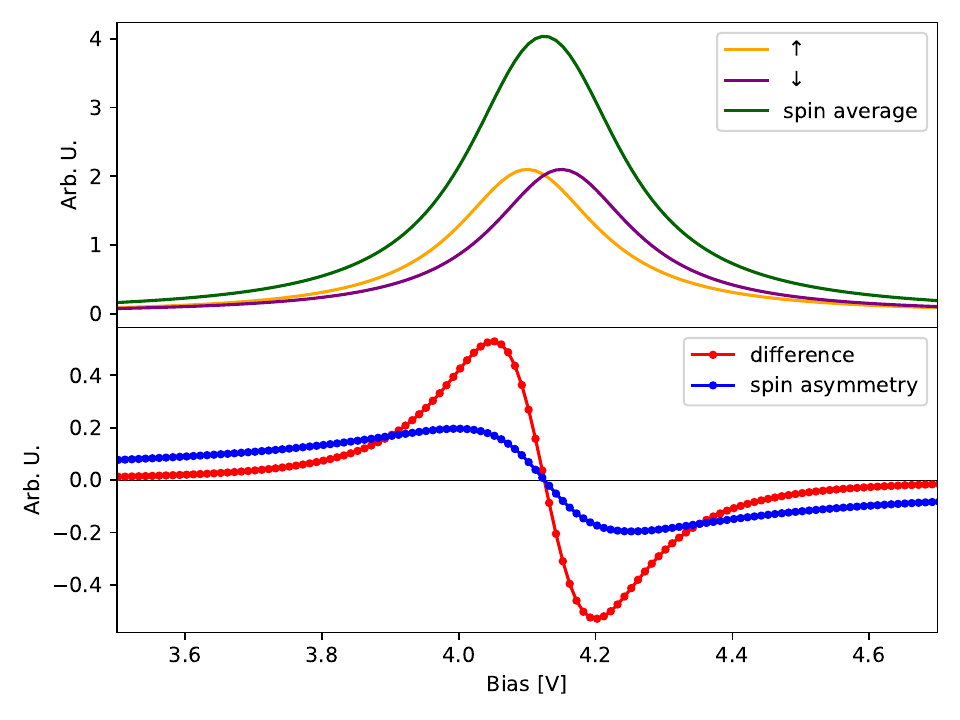}
		\caption{\textbf{Model: Symmetric spin-split peaks in d$I$/d$U(U)$.} 
		\textbf{Top:}~$|\uparrow>$ (orange) and $|\downarrow>$ (purple) states with a spin-split of $50$\,mV. 
		A spin averaged peak (green) as result from a sum of both channels.
		\textbf{Bottom}:~Difference in d$I$/d$U(U)$~(red) and computed spin asymmetry~(blue) between $|\uparrow>$ and $|\downarrow>$.
}
	\label{symetric_lor}
\end{figure}

It has been shown before, that the maximum difference of the spin-dependent d$I$/d$U(U)$ curves is not necessarily located at the center of an IPS peak~\cite{Kubetzka2007}. 
In the following we will introduce a simple model step by step in order to visualise and discuss this experimental observation in more detail. 
Since spin-polarized IPSs are spin-split, we start by introducing two symmetric and equal Lorentzians, one for $|\uparrow>$ and one for $|\downarrow>$ with a spin-split of 50 mV (see Fig.~\ref{symetric_lor}).
The spin averaged signal is a sum of both peaks. 
If these states would be measured consecutively with a $100\,\%$ polarised tip magnetized in opposite directions along the axis of magnetization of the sample the measured signals would look like in the top panel (yellow and purple curve, respectively). The intrinsic splitting would be resolved. 
In the bottom panel, the d$I$/d$U(U)$ difference and spin asymmetry (as defined in Section~\ref{asy_intro}) computed from the spin resolved curves are shown.
Here, it becomes clear that at the maximum of the spin averaged peak we observe a vanishing difference in d$I$/d$U(U)$ and therefore a vanishing spin asymmetry.
The d$I$/d$U(U)$ difference and asymmetry are highest at the rising and falling flank of the spin averaged peak, which corresponds also to the rising and falling flanks of the individual spin-split peaks, respectively.

\subsection{Finite probe spin-polarization}\label{finit_pol}
\begin{figure}[!htb]
	\centering
		\includegraphics[width=0.6\columnwidth]{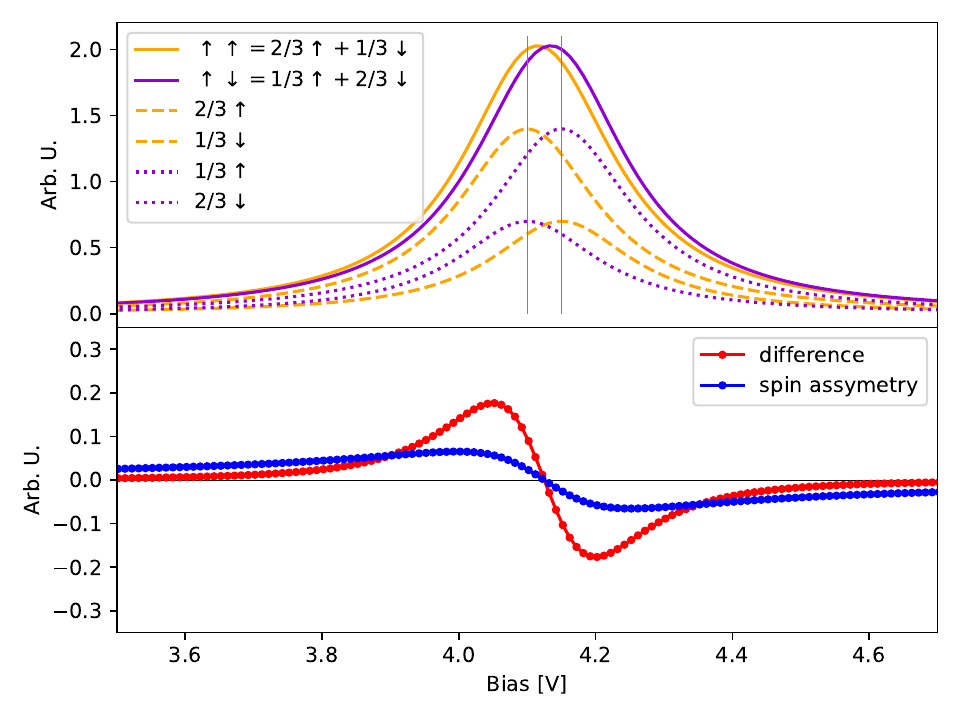}
		\caption{\textbf{Model: Symmetric spin-split peaks in d$I$/d$U(U)$ with finite tip spin-polarization.} 
		\textbf{Top:}~$|\uparrow\uparrow> = 2/3|\uparrow> + 1/3|\downarrow>$ and $|\uparrow\downarrow> = 1/3|\uparrow> + 2/3|\downarrow>$ states with an intrinsic spin-split of 50~meV.
		Orange curves: $|\uparrow>$  and $|\downarrow>$ contributions (dashed) and resulting $|\uparrow\uparrow>$ peak (solid) for a finite tip polarization.
		Violet curves: resulting $|\uparrow\downarrow>$ peak (solid) and its contributions (dotted) for an inverted tip polarization.
		Gray lines mark positions of the intrinsic $|\uparrow>$  and $|\downarrow>$ peaks.
		\textbf{Bottom}:~Difference in d$I$/d$U(U)$~(red) and computed spin asymmetry~(blue) between $|\uparrow\uparrow>$ and $|\uparrow\downarrow>$.
}
	\label{mix_lor}
\end{figure}

The spin-split IPSs are expected to significantly overlap in energy, due to a relatively small spin-split compared to their intrinsic lifetime broadening~\cite{Passek1995}.
As a consequence, at a given energy both states are accessible to the respective spin-polarized tunneling electrons.
Since in SP-STM, magnetic probe tips are usually less then $100\,\%$ spin-polarized, at each energy we observe mixtures of $|\uparrow>$ and $|\downarrow>$ IPSs.
This has already been discussed in Ref.~\cite{Schlenhoff2012} and especially in detail in the supplemental online material of Ref.~\cite{Schlenhoff2019}. 
Assuming a $33\,\%$ spin-polarization of the tip the contributions of the $|\uparrow>$ and $|\downarrow>$ tunneling channels are weighted accordingly, resulting in an spin-dependent peak whose maximum is shifted with respect to the pure $|\uparrow>$ and $|\downarrow>$ peaks towards the spin averaged peak maximum.
Fig.~\ref{mix_lor} shows such a situation with $|\uparrow\uparrow> = 2/3|\uparrow> + 1/3|\downarrow>$ and $|\uparrow\downarrow> = 1/3|\uparrow> + 2/3|\downarrow>$ peaks.
In the top panel the dashed orange lines represent contributions to the $|\uparrow\uparrow>$ state (solid orange line) according to the tip polarization while the dotted violet curves show contributions to the $|\uparrow\downarrow>$ state (solid purple line) upon reversal of the tip magnetization. 
In the bottom panel the difference in d$I$/d$U(U)$ (red) and spin asymmetry (blue) is shown.  
As a result of this convolution, the effective spin-split is much smaller than the intrinsic one~\cite{Schlenhoff2019}, since the maxima of the $|\uparrow\uparrow>$ and $|\uparrow\downarrow>$ peaks are closer to the maximum of the spin averaged signal. 
Note that the evolution of both the d$I$/d$U(U)$ difference and the spin asymmetry looks the same as in the previous case, but with smaller absolut values~(compare Fig.~\ref{symetric_lor}). 

\subsection{Dependence of d$I$/d$U(U)$ difference and spin asymmetry on d$I$/d$U(U)$ signal intensity}\label{assy_win}
\begin{figure}[!htb]
	\centering
		\includegraphics[width=0.6\columnwidth]{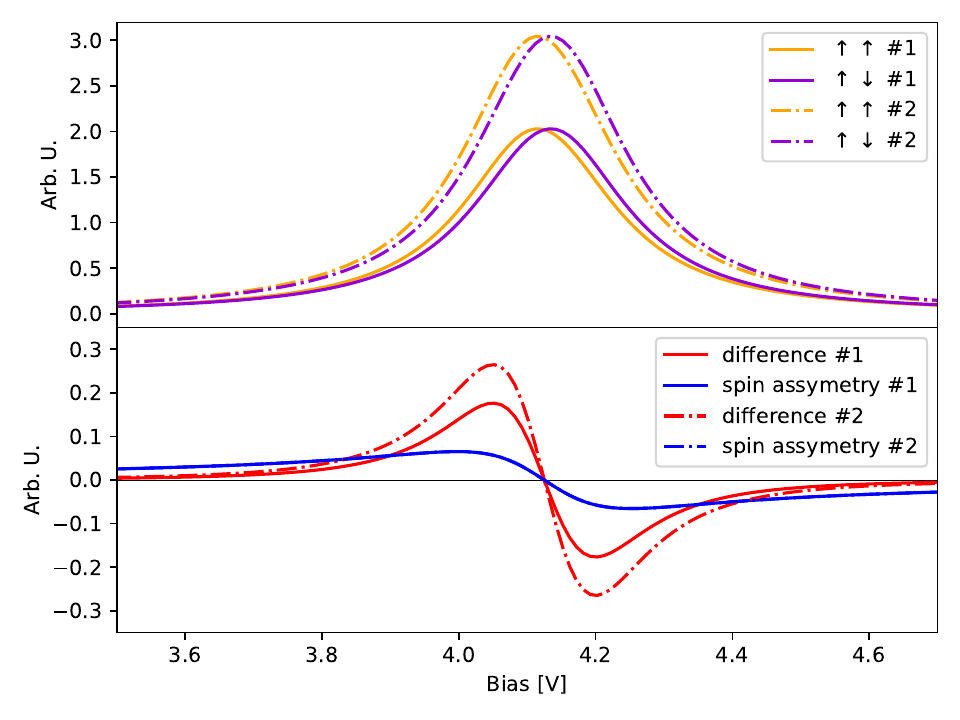}
		\caption{\textbf{Model: Dependence of d$I$/d$U(U)$ difference and spin asymmetry on d$I$/d$U(U)$ signal intensity.} 
		\textbf{Top:}~Two pairs of mixed $|\uparrow\uparrow>$ (orange) and $|\uparrow\downarrow>$ (purple) states with an intrinsic spin-split of 50~meV. 
		Series $\#1$ (solid lines) is identical to the peaks shown in Fig.~\ref{mix_lor}. 
		Series $\#2$ (dash-dot) has a by $50\,\%$ higher amplitude, but is composed of the same proportions of $|\uparrow>$ and $|\downarrow>$ contributions as the series $\#1$. 
		\textbf{Bottom}:~Difference in d$I$/d$U(U)$ (red) and computed spin asymmetry (red) between $|\uparrow\uparrow>$ and $|\uparrow\downarrow>$ for both series.
		While the d$I$/d$U(U)$ difference of series $\#2$ is increased compared to that of series $\#1$, the spin asymmetry is unaffected.}
	\label{assywin}
\end{figure}

We can use this very simple case to demonstrate the superiority of the spin asymmetry signal over the d$I$/d$U(U)$ difference for determining the spin-polarization. 
In Fig.~\ref{assywin}, the $|\uparrow\uparrow>$ and $|\uparrow\downarrow>$ peaks from Fig.~\ref{mix_lor}, labeled by $\#1$, are shown again.
In addition, two exemplary curves ($\#2$) are shown, being composed of the same weighted contributions $|\uparrow\uparrow> = 2/3|\uparrow> + 1/3|\downarrow>$ and $|\uparrow\downarrow> = 1/3|\uparrow> + 2/3|\downarrow>$, respectively, as before, but with a by $50\,\%$ increased height of the respective intrinsic peaks $|\uparrow>$ and $|\downarrow>$.  
Thus, the peaks $|\uparrow\uparrow>$ $\#1$ and $|\uparrow\uparrow>$ $\#2$ (and accordingly their reversed configurations $|\uparrow\downarrow>$ $\#1$ and $|\uparrow\downarrow>$ $\#2$) have the same spin polarization but exhibit different signal intensities. 
As a result, in the bottom panel of Fig.~\ref{assywin} we can see that the d$I$/d$U(U)$ difference signal for the case~$\#2$ has also increased by $50\,\%$. 
However, according to the underlying unchanged spin polarization, the spin asymmetry remains unchanged. 
The two spin asymmetry curves for the case $\#1$ and $\#2$, respectively, fall on top of each other.
Consequently, when considering the spin-polarization of states, only the spin asymmetry signal is a robust and reliable measure~\cite{Wortmann2001, Kubetzka2003, Wiesendanger2009}. 

\subsection{Asymmetric shape}\label{assymetric_shape}
\begin{figure}[!htb]
	\centering
		\includegraphics[width=0.6\columnwidth]{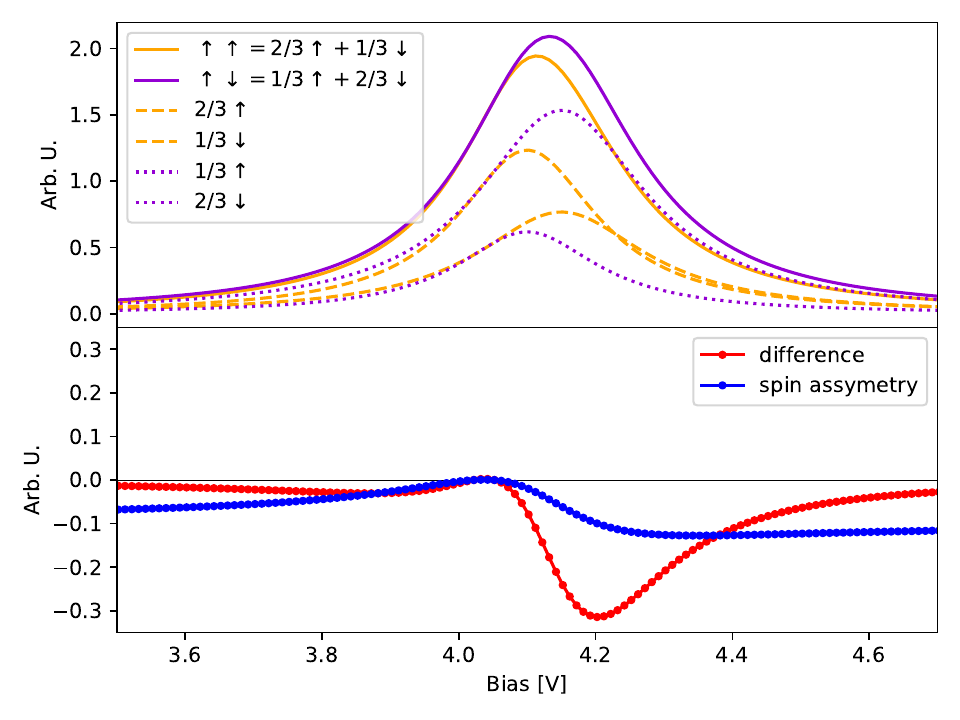}
		\caption{\textbf{Model step I:  Asymmetric spin-split peaks in d$I$/d$U(U)$ - different heights and widths.} 
		\textbf{Top:}~$|\uparrow\uparrow> = 2/3|\uparrow> + 1/3|\downarrow>$ and $|\uparrow\downarrow> = 1/3|\uparrow> + 2/3|\downarrow>$ states with an intrinsic spin-split of 50~meV.
		Orange curves: $|\uparrow>$  and $|\downarrow>$ contributions (dashed) and resulting $|\uparrow\uparrow>$ peak (solid) for a finite tip polarization.
		Violet curves: resulting $|\uparrow\downarrow>$ peak (solid) and its contributions (dotted) for an inverted tip polarization.
		Here, the intrinsic $|\downarrow>$ state has an increased amplitude and width compared to the  $|\uparrow>$  state.
		As a result, the purple solid curve $|\uparrow\downarrow>$ is wider and has a larger amplitude.
		\textbf{Bottom}:~Difference in d$I$/d$U(U)$ (red) and computed spin asymmetry (blue) between $|\uparrow\uparrow>$ and $|\uparrow\downarrow>$.
	}
	\label{nonequal}
\end{figure}

\begin{figure}[!htb]
	\centering
		\includegraphics[width=0.6\columnwidth]{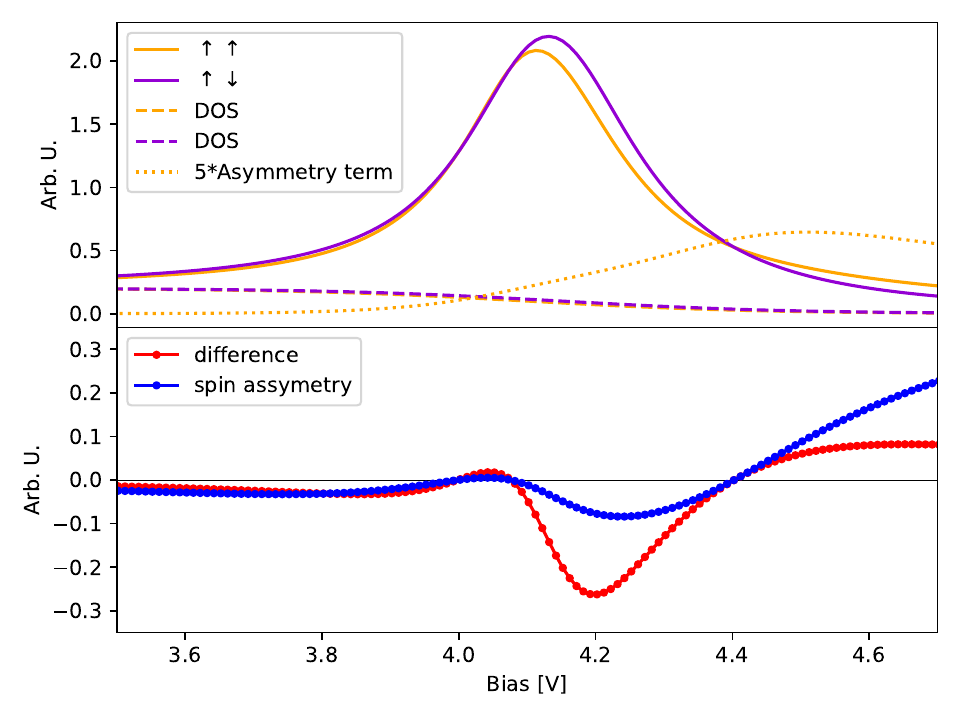}
		\caption{\textbf{Model step II:  Asymmetric spin-split peaks in d$I$/d$U(U)$ - different falling flanks.} 
		\textbf{Top:}~$|\uparrow\uparrow> = 2/3|\uparrow> + 1/3|\downarrow>$ (orange, solid) and $|\uparrow\downarrow> = 1/3|\uparrow> + 2/3|\downarrow>$ (purple, solid) peaks with an intrinsic spin-split of 50~mV, including sample DOS contributions (dashed orange and violet, respectively) and an asymmetric term (orange, dotted). 
		(The latter has been enlarged by a factor of 5 for a better visibility.) 
		\textbf{Bottom}:~Difference in d$I$/d$U(U)$ (red) and computed spin asymmetry (blue) between $|\uparrow\uparrow>$ and $|\uparrow\downarrow>$ peaks.
		}
	\label{asymmetric_mix}
\end{figure}

In our experiments, the spin-dependent IPS peaks are asymmetric and exhibit slightly different shapes with respect to each other. 
Concrete, they exhibit slightly different heights, widths, and they also have different falling flanks (see Fig.~4 in the main text).
This results in an evolution of the spin asymmetry signal as a function of bias, that is much more complex then the cases discussed so far.
We include these experimental observations in our model in two steps. 
First, we adjust the amplitude and width of the peaks such that the $|\uparrow\downarrow>$ would dominate the spectroscopy at all energies.
The intrinsic $|\downarrow>$ state is hereby modeled by a a larger amplitude and a broader linewidth, affecting the resulting mixed $|\uparrow\uparrow>$ and $|\uparrow\downarrow>$ peaks according to its respectively different weighted contribution.
The result can be seen in Fig.~\ref{nonequal}. 
The spin asymmetry becomes negative in the whole energy range around the state. 
In our experimental data (see Fig.~4), however, the spin asymmetry changes sign several times. 
Hence, to get closer to our experimental data we need to include two more factors. 
First, we include a non-vanishing DOS for bias voltages lower than the first IPS.
Starting from a low bias, where direct tunneling into the surface is still possible, we assume a constant sample DOS to which electrons can tunnel from the tip during the bias sweep until the first IPS is reached.
At this point, the tip retracts more then $200$\,pm (in total more than $500$\,pm compared to tunneling at 1~V), as deduced from our measured $z(U)$ characteristic (see. Fig.~\ref{combination_top}), and the probability of tunneling directly into the sample DOS is close to zero. 
We model this contribution of direct tunneling at the low-energy side of the first IPS with a $B*(1-tanh(U-E_{1st}))$ function that is placed at the maximum of the first low-energy $|\uparrow>$ peak, $E_{1st}$, where its value drops from $2*B$ to zero, when sweeping the bias $U$.
Second, we include an asymmetric term which accounts for inelastic processes mediated over the nearest lower-in-energy IPS.
In order to implement this, we modify the model by adding a $A*(1+tanh(\Gamma * (U-E_{1st}-\Gamma)))$ step located at the intrinsic low energy $|\uparrow>$ peak position $E_{1st}$ with an amplitude $A$ on the order of $1-5$\,$\%$ of the peak's amplitude, where $\Gamma$ is the broadening parameter. Following the idea from Ref.~\cite{Doniach1970} we multiply this term by $1 - \int\limits_{0}^{\infty} \frac{\gamma^2}{(U-E_{1st})^2 + \gamma^2}dE$ which ensures that the asymmetric term is only viable in proximity of the nearest lower-in-energy state.
Here, $\gamma$ is the half-width of the IPS state.
This term induces an asymmetric shape of the peak curves, resulting in different falling flanks for $|\uparrow\uparrow>$ and $|\uparrow\downarrow>$ configuration. 
Fig.~\ref{asymmetric_mix} shows the final model of a pair of $|\uparrow\uparrow>$ and $|\uparrow\downarrow>$ peaks. 
Compared to Fig.~\ref{nonequal}, it includes now the additional DOS contributions and the asymmetry term, that are both plotted as well in the top panel. 
As a result, the $|\uparrow\downarrow>$ curve is higher at lower bias due to its larger amplitude and width, as it has already been in the previous step. 
The sample DOS contribution being tight to the intrinsic positions of the $|\uparrow>$ and $|\downarrow>$ affects the spin asymmetry only marginally around the state center.
While in the previous Fig.~\ref{nonequal}, the $|\uparrow\downarrow>$ curve runs continuously above the $|\uparrow\uparrow>$ curve, the inclusion of the asymmetric term (magnified by a factor of $5$ in Fig.~\ref{asymmetric_mix}), results in two crossings of the $|\uparrow\uparrow>$ curve with the $|\uparrow\downarrow>$ curve, running then above the latter.  
One crossing point is located on the rising flanks of the IPS peaks and the second is found on the bottom of the falling flanks of the IPS peaks.

Compared to the fully symmetric starting model (see Fig.~\ref{symetric_lor}), the spin asymmetry, shown in the lower panel of  Fig.~\ref{asymmetric_mix}, becomes much smaller on the rising flank and its first local maximum is shifted towards the maximum of the IPS state. 
The zero crossing of the spin asymmetry from Fig~\ref{mix_lor} is no longer located in between the peaks' maxima but on the rising flanks. 
Furthermore, the extremes of the spin asymmetry do not coincide with the extremes of the d$I$/d$U(U)$ difference. 
While the extremes of the difference in d$I$/d$U(U)$ remains at the same position over all steps introduced in our model, the maximum absolute value of spin asymmetry close to the IPS center changes its position.
Most significantly, the asymmetric term enables the experimentally observed change in sign of the spin asymmetry at the falling flanks of the peaks. 
Here, since the difference in d$I$/d$U(U)$ is significant above $4.5$\,V in Fig.~\ref{asymmetric_mix} and the overall signal drops, the spin asymmetry increases, consistent with the predominance of one spin channel at the long tail. 

\subsection{Two IPSs spin-split model}\label{model_full_text}
In real measurements, the next higher-in-energy spin-split pair of peaks already arises where signals from the previous spin-split pairs are small. 
To demonstrate this, we include a second pair of Lorentzians modified in a similar fashion and shifted to $5.4$\,V.
\begin{figure}[!htb]
	\centering
		\includegraphics[width=0.6\columnwidth]{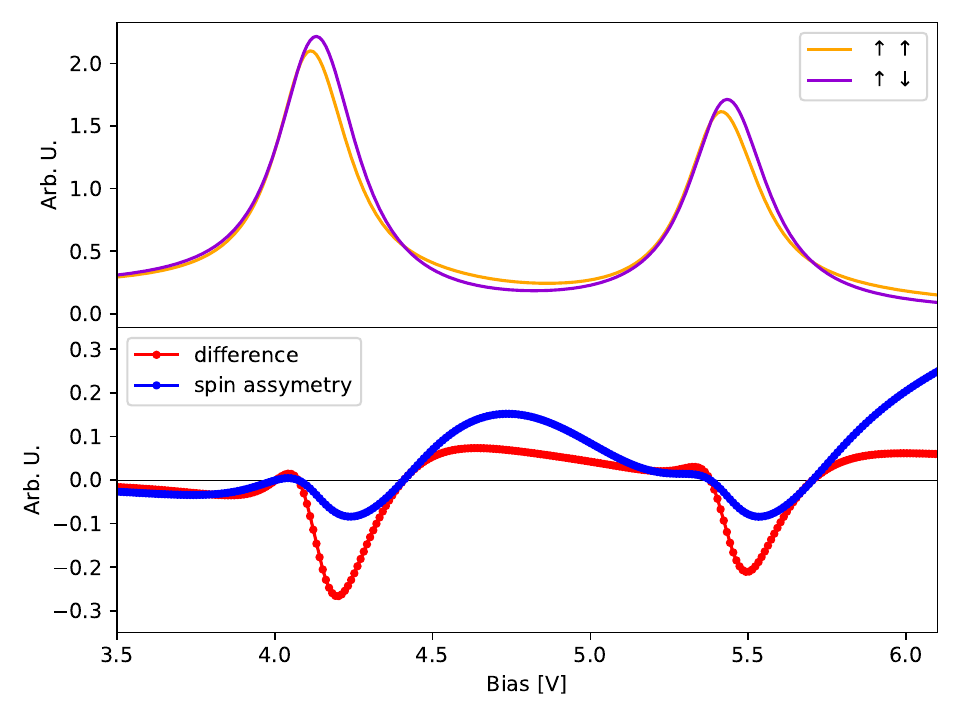}
		\caption{\textbf{Model: Adjacent pairs of asymmetric spin-split peaks in d$I$/d$U(U)$.} 
		\textbf{Top:}~An effective $|\uparrow\uparrow>$ curve as a sum of two pairs of asymmetric $|\uparrow\uparrow>$ states in orange. 
		An effective $|\uparrow\downarrow>$ curve as a sum of two pairs of asymmetric $|\uparrow\downarrow>$ states in purple. The intrinsic spin-split is set to 50~mV.
		\textbf{Bottom}:~Difference in d$I$/d$U(U)$ (red) and computed spin asymmetry (blue) between $|\uparrow\uparrow>$ and $|\uparrow\downarrow>$ curves.
}
	\label{model_full}
\end{figure}

Fig.~\ref{model_full} shows a simple model of spin-split, asymmetric-in-shape $n=1$ and $n=2$ IPS peaks detected in spin-resolved d$I$/d$U(U)$ spectroscopy with a finite spin polarization of the tip as described above.  
A positive asymmetry is found on the rising flanks and in-between the peaks, while a negative spin asymmetry is found only on the falling flanks of the peaks. 
In close proximity of the spin-split peak maxima, the spin asymmetry behaves in a well understood way~\cite{Kubetzka2007,Schlenhoff2012, Schlenhoff2019}.
Here, due to the spin-dependent energetic shift of the peaks, the $|\uparrow\uparrow>$ curve runs above the $|\uparrow\downarrow>$ curve on the rising flank, while the situation reverts on the falling flank.
At the high-energy tail of the peaks, their asymmetric shape dominates the spin asymmetry.
Note that in the model, the highest spin polarization is achieved at the long tail of the spin-split peaks due to a spin selective inelastic channel mediated over the nearest lower-in-energy $|\uparrow>$ IPS (see Section~\ref{assymetric_shape}).

In conclusion, the model demonstrates how a considerable difference in the spin-dependent d$I$/d$U(U)$ signal and concomitant a considerable spin asymmetry can arise between adjacent spin-split peaks due to their spin dependent asymmetric shape.
In the following, we will use this model for the discussion of our spin-resolved spectroscopy data. 

\newpage
\section{Energy dependence of spin-dependent signals in spin-resolved IPS spectroscopy on Gr/Fe/Ir(111)}\label{enerdy_dependence}

\begin{figure}[!htb]
	\centering
		\includegraphics[width=0.8\columnwidth]{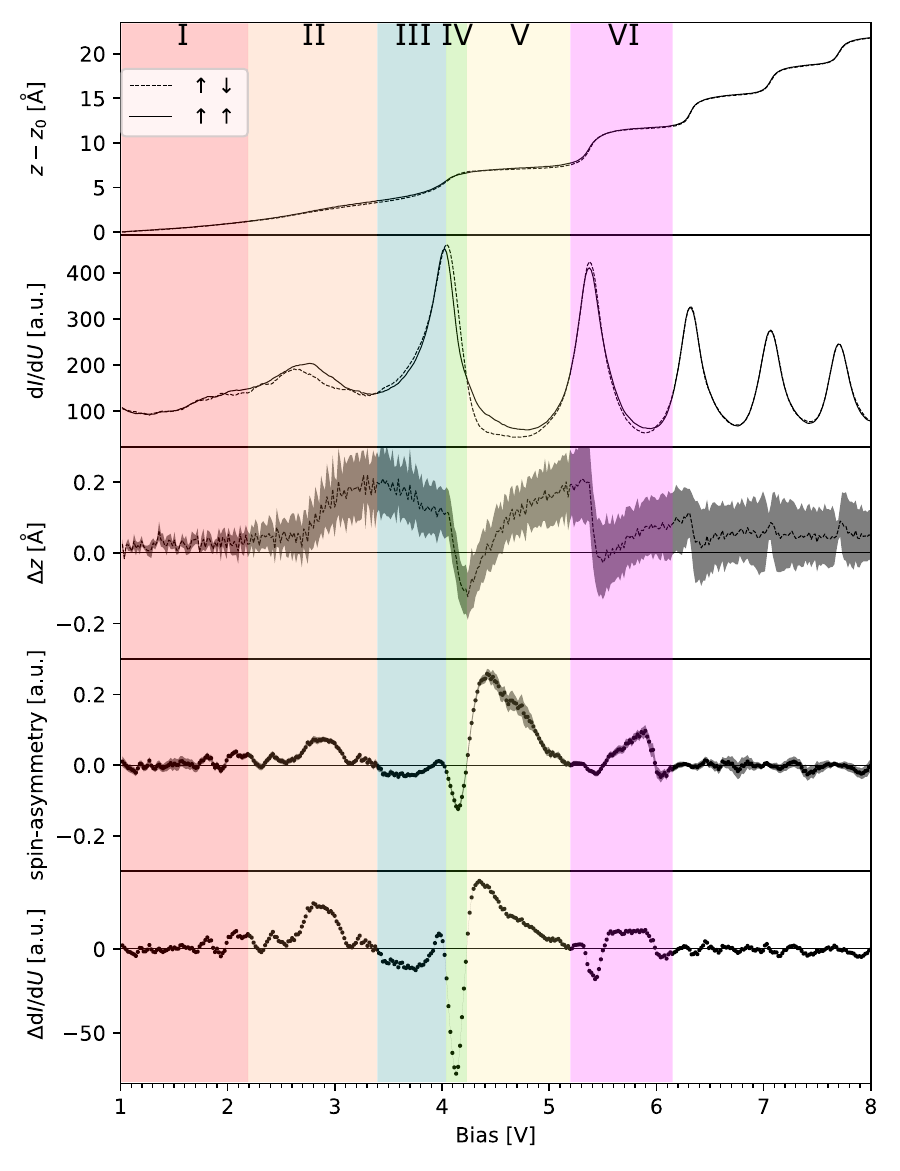}
		\caption{\textbf{Spin-dependent signals in constant current SP-STS of IPSs on the moiré top site.} 
		Spin-polarized constant current tunneling spectra on the top site of the Gr/Fe/Ir(111) moiré structure.
		From the top: $(z(U)-z_0)$, d$I$/d$U(U)$, $\Delta z$, spin asymmetry and $\Delta$d$I$/d$U(U)$. 
		Stabilization parameters: $U=1$\,V, $I_{\rm t}=1$\,nA.
        Spectra were taken with Fe coated W tip in external magnetic field of $\pm1$~T applied along the tip axis.
        Solid lines represents $|\uparrow\uparrow>$ configuration, while dotted lines represents $|\uparrow\downarrow>$ configuration.
        Each curve represents an average of 3 consecutive curves. 
        Shade on $\Delta z$ is the error calculated as a standard deviation of averaged curves for $|\uparrow\uparrow>$ and $|\uparrow\downarrow>$ configurations.
        Shade on spin asymmetry is the error calculated as asymmetry of the initial $|\uparrow\downarrow>$ and control spectra after the whole series. 
        Different regimes marked with color shadings and labelled by I-IV.}
	\label{combination_top}
\end{figure}

\subsection{Data presentation}\label{data}
\begin{figure}[!htb]
	\centering
		\includegraphics[width=0.8\columnwidth]{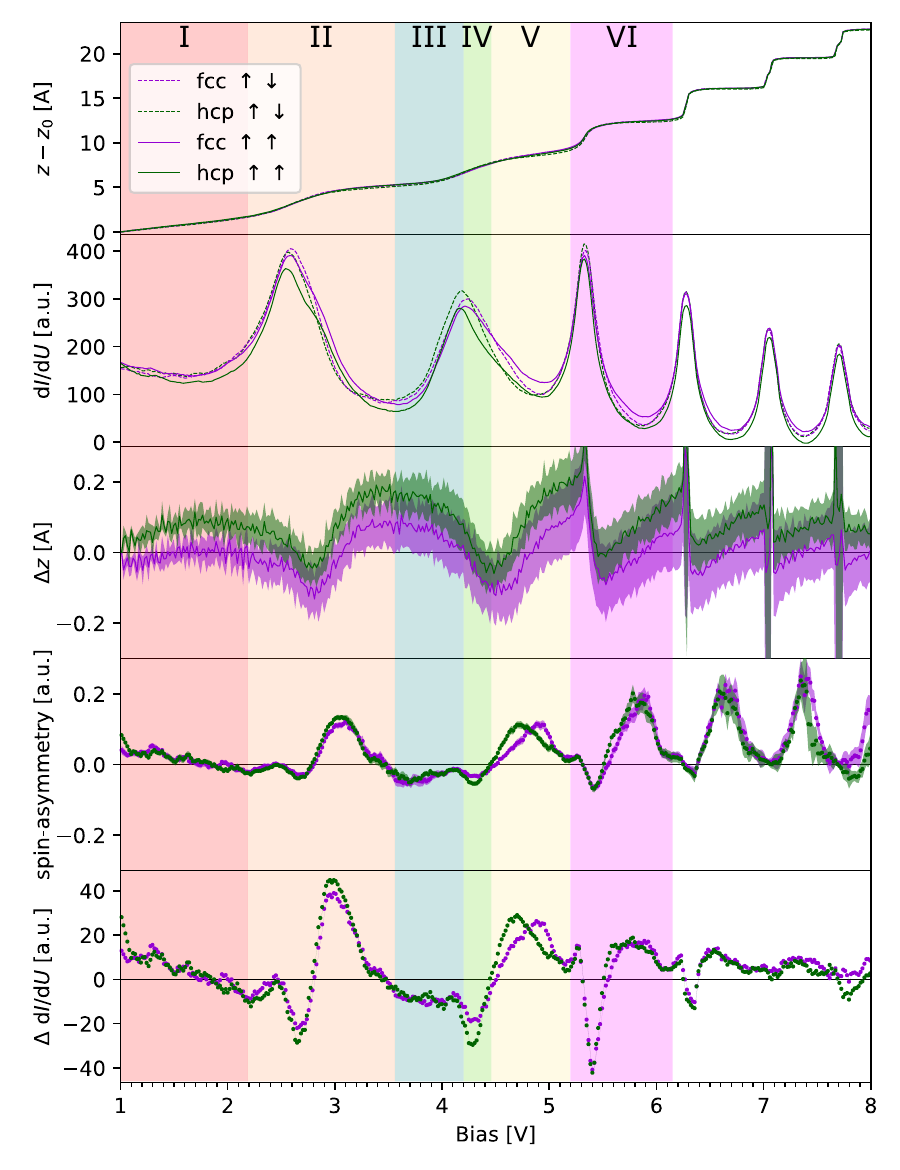}
		\caption{\textbf{Spin-dependent signals in constant current SP-STS of IPSs on the moiré valleys.} 
		Spin-polarized constant current tunneling spectra on fcc and hcp position of the Gr/Fe/Ir(111) moiré structure. 
		From the top $(z(U)-z_0)$, d$I$/d$U(U)$, 
        $\Delta z$, spin asymmetry and $\Delta$d$I$/d$U(U)$. Stabilization parameters: $U=1$\,V, $I_{\rm t}=1$\,nA.
        Spectra were taken with Fe coated W tip in external magnetic field of $\pm1$~T applied along the tip axis.
        Solid lines represents $|\uparrow\uparrow>$ configuration, while dotted lines represents $|\uparrow\downarrow>$ configuration.
         Each curve represents an average of 3 consecutive curves. 
        Shade on $\Delta z$ is the error calculated as a standard deviation of averaged curves for $|\uparrow\uparrow>$ and $|\uparrow\downarrow>$ configurations.
        Shade on spin asymmetry is the error calculated as asymmetry of the initial $|\uparrow\downarrow>$ and control spectra after the whole series.  
        Different regimes marked with color shadings and labelled by I-IV.
        }
        \label{combination_fcchcp_1}
\end{figure}

When performing constant current spectroscopy of IPSs, two different channels are recorded simultaneously while ramping the sample bias $U$. 
In the so-called topography channel, we record the apparent height $z(U)$, which in this specific case is a measure of how far the tip retracts during the bias sweep (see Section~\ref{asy_intro}).  
The second recorded channel is the differential conductance d$I$/d$U(U)$ which is a measure of the DOS at a given energy.
We stabilize the tip-sample distance at a bias such that the integrated spin-polarized current from the Fermi level to that bias vanishes (see Section~\ref{asy_intro}), which is in this particular case at $U=1$\,V. 
As a consequence, we have the same absolut distance $z_0$ at $U=1$\,V in both spin-resolved measurement configurations, once for parallel and once for antiparallel alignment of tip and sample magnetic moments.  
Therefore, we can plot the apparent height $(z-z_0)$ which allows for comparison between measurements in different tip-sample magnetic configurations.
$\Delta z$ is here defined as the difference between the two spin-dependent $(z-z_0)$ curves. This spin-dependent 'corrugation' in the constant current mode is a measure for the spin-dependent difference in the current, that integrates up to the energy determined by the applied bias $U$. 
Hence, it reflects the accumulated spin-dependent signal in the apparent height. 
From the two measured configurations of d$I$/d$U(U)$ we can compute the difference, $\Delta$d$I$/d$U(U)$, and the spin asymmetry.
The first is proportional to the difference in DOS of spin up and spin down tunneling channels, the latter is this difference normalized by the spin averaged signal and therefore a measure of the spin polarization at a given energy e$U$ (see Section~\ref{asy_intro}).
We will now analyze apparent height, $\Delta z$, d$I$/d$U(U)$, $\Delta$d$I$/d$U(U)$ and spin asymmetry in case of the moiré top position of the Fe intercalated Gr system step by step.
The data presented in Fig.~\ref{combination_top} are part of the data set shown in Fig.~4 of the main text.
The presented analysis can be then repeated analogously for the fcc and hcp positions. 
(The respective data are shown in Fig.~\ref{combination_fcchcp_1}, and to some part in Fig.~4 of the main manuscript.) 
For clarity we split the spectra into six regions:
\begin{enumerate}[label=\Roman*.]
\item without any spectral features,
\item comprising the interface state (IFS), 
\item the rising flank of the 1st IPS peak, 
\item the falling flank of the 1st IPS peak, 
\item in-between peaks, 
\item 2nd IPS peak and the energy above it.
\end{enumerate}
Region I is characterized by a flat featureless DOS that exhibits an insignificant spin polarization, as can be deduced from the $\Delta z$, $\Delta$d$I$/d$U(U)$ and spin-asymmetry values, all being almost zero.
In the transition between regions I and II small differences in d$I$/d$U(U)$ are observed with the $|\uparrow\uparrow>$ curve being slightly higher.
Those are indications of a spin-polarized unoccupied sample DOS stemming from $d$-states that reach comparably far into the vacuum~\cite{BazarnikGdAu}.
This mostly positive spin-asymmetry leads to a small rise in $\Delta z$.

When we move to region II, the IFS arise in its centre. 
While the IFS is predominantly located on the fcc and hcp sites at these energies (compare Fig.~\ref{combination_fcchcp_1} and Fig.~4), a faint signature also appears on the top site. 
The $|\uparrow\uparrow>$ configuration has its maximum shifted towards higher energies by 0.15~V which leads to a significant $\Delta$d$I$/d$U(U)$ and spin-asymmetry, respectively, and in consequence to a rising $\Delta z$ on the falling flank of this peak.

Above the IFS both d$I$/d$U(U)$ curves converge and cross at $U=3.4$\,V entering the region III, at which point $\Delta z$ reaches a local maximum of $0.2$~\AA.
The following regions III-VI are well described by our model. 
At the bottom of the rising flanks of the $n=1$ IPS peaks the $|\uparrow\downarrow>$ d$I$/d$U(U)$ curve runs above the $|\uparrow\uparrow>$ curve, leading to a negative spin-asymmetry and thus to a small drop in $\Delta z$, which still remains significant at around $0.1$~\AA. 
The d$I$/d$U(U)$ curves cross close to 3.9~V, and a small local maximum in the spin-asymmetry is visible close to the $n=1$ peak maxima. 
This is what we would expect for a spin-split state as described by our model (see Section~\ref{assy_section}).
As in the model, the $|\uparrow\uparrow>$ peak is slightly lower, and even at its maximum it runs below the $|\uparrow\downarrow>$ peak, which is higher and shifted towards higher energies by $15$\,mV. 
As deduced from our model, the observed difference in width and amplitude indicates a spin-dependent probability density and lifetime of the spin-split IPS~\cite{Passek1995, Nekovee1993} for the here studied complex Gr-ferromagnet hybrid system.
The $\Delta$d$I$/d$U(U)$ and spin-asymmetry change sign already at the maximum of the $|\uparrow\uparrow>$ d$I$/d$U(U)$ peak, exhibiting a very sharp minimum covering the falling flanks of the $n=1$ IPSs, as observable in region~IV.
The large $\Delta$d$I$/d$U(U)$ leads to a sudden drop and change of sign in $\Delta z$ at $U=4.12$\,V.
The $\Delta z$ then reaches its most negative value of $-0.125$\,\AA\, at $U=4.225$\,V, where the $\Delta$d$I$/d$U(U)$ and spin-asymmetry change sign yet again due to the highly asymmetric shape of the $|\uparrow\downarrow>$ d$I$/d$U(U)$ peak, as discussed above.

As in our model, the whole region V is dominated by the influence of the asymmetric shape of both $n=1$ peaks.
This leads to highly spin dependent signals: the high $\Delta$d$I$/d$U(U)$ and spin-asymmetry cause rise in the $\Delta z$ which crosses zero at $U=$4.4\,V and continues increasing to the end of region V where the spin-asymmetry vanishes at $5.2$\,V.

In region VI on the rising flanks of the spin-split peaks we observe a very small $\Delta$d$I$/d$U(U)$ and spin-asymmetry due an overlap with the long-tail of the low-energy $|\uparrow\uparrow>$ peak (compare Fig.~\ref{model_full}).
As in our model, again the $n=2$ $|\uparrow\uparrow>$ peak is lower in height, and therefore, the $\Delta$d$I$/d$U(U)$ and spin-asymmetry change sign already in front of the maximum of the spin-split peaks (see Sections~\ref{assymetric_shape} and~\ref{model_full_text}). 
Here, the higher-energy $|\uparrow\downarrow>$ d$I$/d$U(U)$ peak dominates causing a negative $\Delta$d$I$/d$U(U)$ and spin-asymmetry provoking also a sharp drop in $\Delta z$.
Due to the asymmetric shape of the spin-split peaks, being less pronounced for $n=2$, the d$I$/d$U(U)$ cross at half peak height close to $U=5.5$\,V leading to a positive spin-asymmetry and an increase in $\Delta z$ again. 

For the $n>2$ IPSs on each moiré site, a spin-splitting of the peaks in d$I$/d$U(U)$ cannot be resolved anymore, indicating a decrease of the spin-splitting below the set energy resolution of the spectra with increasing~$n$ (see Fig.~\ref{combination_top} and \ref{combination_fcchcp_1}). 
This is generally expected from the increasing delocalization of the IPS wavefunctions perpendicular to the surface with increasing order~$n$ and the concomitant decrease in overlap with a spin-polarized surface~\cite{Donath2007}. 
However, especially on the valleys, the d$I$/d$U(U)$ curves in this high energy range still exhibit a spin-dependence in terms of slightly different shapes (e.g. height, width, tail), resulting in pronounced spin asymmetry maxima in the regions where the long tails of the $(n-1)$ IPS peaks overlaps with the rising flanks of the $n$ IPS peaks. 

\subsection{Interpretation}

To conclude, the energy dependent evolution of the spin-dependent signals in spin-resolved constant current spectroscopy of IPSs on different positions within the moiré unit cell of the Gr-Fe hybrid can be well-understood as a result of spin-polarized IPSs that mediate the spin-dependent resonant electron tunneling process from the tip to the surface. 
In section~\ref{data}, the two simultaneously recorded channels, $z(U)$ and d$I$/d$U(U)$, recorded for parallel and antiparallel configuration of tip and sample magnetization, respectively, and derived spin-dependent signals (e.g. $\Delta z$, $\Delta$d$I$/d$U(U)$ and spin asymmetry) are brought into a meaningful relationship to each other. We have developed a simple model in Section~\ref{assy_section} that reproduces the prominent features of our spin-resolved IPS spectroscopy data. 

Note, that slightly different heights and widths of overlapping spin-split IPS peaks have been found before in SP-STS~\cite{Kubetzka2007,Schlenhoff2012}. 
However, to such an extent asymmetrically shaped spin-split IPS peaks have not been observed on the yet studied much simpler, pure thin-film metal systems~\cite{Kubetzka2007,Schlenhoff2012, Schlenhoff2019}.
Pascual~\textit{et al.}~\cite{Pascual2007} were the first that discussed an asymmetric shape of IPSs in (spin-averaged) constant current spectroscopy. 
They attributed the asymmetry to the 'dynamic' measurement mode, where the tip retracts from the surface while tunneling into the IPS, thus leading to some peak shape distortion. 
In their model calculation of this 'dynamic' situation, they show a very sharp falling flank, similarly shaped as observed in our experiments e.g. for the moiré top $n=1$  $|\uparrow\downarrow>$ peak. 
However, this simulated drastic asymmetry has not been observed in their experiment, and the additional broadening has been assigned to inelastic processes. 
In our model, we include an additional inelastic tunneling channel mediated over the nearest lower-in-energy $|\uparrow>$ IPS to explain the different shapes of $|\uparrow\downarrow>$ and $|\uparrow\uparrow>$ spectroscopic curves. 
Since we find in our experiments that this asymmetric peak shape is spin dependent, we expect this inelastic tunneling channel to depend on the electron spin orientation.
Moreover, since we have never observed this behavior in an all metal system like Fe/W(110), Fe/Ir(111) or Co/Ir(111), we deduce that the presence of Gr on the ferromagnetic metal system enables or enhances this spin-dependent excitation process.
However, the nature of this Gr-induced spin-dependent excitation cannot be determined without more precise calculations and is beyond the scope of this work.